\documentclass{birkjour}
 \theoremstyle{definition}
 
 \theoremstyle{remark}

 \numberwithin{equation}{section}
\usepackage{bm}

\usepackage[utf8]{inputenc}

\usepackage{color,upgreek}
\usepackage{cite}

\usepackage{bm}
\newcommand{\gdual}[1]{\overset{\:{}^{{}^{\boldsymbol{\sim}}}}{\smash[t]{#1}}} 
\newcommand{\dual}[1]{\overset{{}^{{}^{\boldsymbol{\sim}}}}{\smash[t]{#1}}} 
\newcommand{\gdualn}[1]{\overset{\:{}^{{}^{\boldsymbol{\neg}}}}{\smash[t]{#1}}} 
\newcommand{\dualn}[1]{\overset{{}^{{}^{\boldsymbol{\neg}}}}{\smash[t]{#1}}} 

\def\0{\mbox{\boldmath$\displaystyle\mathbb{O}$}}
\def\C{\mbox{\boldmath$\displaystyle\mathbb{C}$}}

\def\J{\mbox{\boldmath$\displaystyle\boldsymbol{J}$}}
\def\vp{\mbox{\boldmath$\displaystyle\boldsymbol{\varphi}$}}
\def\K{\mbox{\boldmath$\displaystyle\boldsymbol{K}$}}
\def\L{\mbox{\boldmath$\displaystyle\boldsymbol{L}$}}
\def\x{\mbox{\boldmath$\displaystyle\boldsymbol{x}$}}
\def\kb{\mbox{\boldmath$\displaystyle\boldsymbol{\kappa}$}}
\def\bz{\mbox{\boldmath$\displaystyle\boldsymbol{\zeta}$}}

\def\bmfj{\mbox{\boldmath$\displaystyle\boldsymbol{\mathfrak{J}}$}}
\def\bmfk{\mbox{\boldmath$\displaystyle\boldsymbol{\mathfrak{K}}$}}

\def\I{\openone}
\def\s{\mbox{\boldmath$\displaystyle\boldsymbol{\sigma}$}}
\def\p{\mbox{\boldmath$\displaystyle\boldsymbol{p}$}}
\def\e{\rm e}

\def\openone{\mathbb I}

\begin{document}

%
%
%
%
%
%
%
%
%

\title[The theory of local mass dimension one fermions]
 {The theory of local mass dimension one \\fermions of spin one half}

\author[Dharam Vir Ahluwalia]{Dharam Vir Ahluwalia}

\address{%
Department of Physics\\
Indian Institute of Technology Guwahati\\
Guwahati 781 039\\ 
India \\}

\address{%
Inter-University Centre for Astronomy and Astrophysics\\
Post Bag 4, Ganeshkhind\\
Pune 411 007\\
India\\ \\}

\address{%
Theoretical Physics Division\\
Physical Research Laboratory\\
Ahmedabad 380 009\\
India\\}

\address{%
Centre for the Studies of the Glass Bead Game\\
Chaugon, Bir\\
Himachal Pradesh 176 077\\
India\\}

\email{d.v.ahluwalia@iitg.ernet.in}



\date{January 1, 2004}
\dedicatory{In memory of papaji, Shri Bikram Singh Ahluwalia}

\begin{abstract}
About a decade ago the present author in collaboration with Daniel Grumiller presented an `unexpected theoretical discovery' of spin one-half fermions with mass dimension one~\cite{Ahluwalia:2004sz,Ahluwalia:2004ab}. In the decade that followed a significant number of groups
explored intriguing mathematical and physical properties of the new construct. However, the formalism suffered from  two troubling features, that of non-locality and a subtle violation of Lorentz symmetry. Here, 
we trace the origin of both of these issues to a hidden freedom in the definition of duals of spinors and the associated field adjoints. In the process, for the first time, we provide a quantum theory of spin one-half fermions that is free from all the mentioned issues. The interactions of the new fermions are restricted to  dimension-four quartic self interaction, and also to a dimension-four coupling with the Higgs. A generalised Yukawa coupling of the new fermions with neutrinos provides an hitherto unsuspected source of lepton-number violation.
The new fermions thus present a first-principle dark matter partner to Dirac fermions of the standard model of high energy physics with contrasting mass dimensions -- that of three halves for the latter versus one of the former without mutating the statistics from fermionic to bosonic.
\end{abstract}

\maketitle

\section{Setting the stage}

Arguing for `some incompleteness' in the earlier works of  Darwin and Pauli,  Dirac in his famous 1928 paper confronts a `duplexity' phenomena, a discrepancy in the observed number of stationary states of an electron in an atom being twice the number given by the then-existing observations and theory~\cite{Dirac:1928hu,Darwin:1927du,Pauli:1927}. He discovers the incompleteness in the previous works and expresses it 
as, ``lying in their disagreement with relativity, or alternatively, with the general transformation theory of quantum mechanics.''
The  solution he proposed, with the subsequent development of the theory of  quantum 
fields~\cite{Tomonaga:1946zz,PhysRev.76.749,PhysRev.75.1736,PhysRev.82.914,Weinberg:1964cn,tHooft:1973mfk,Weinberg:1995mt}, not only resolved the discrepancy but it also introduced a new unexpected duplexity. For each spin one half particle, the theory predicted an antiparticle. Associated with this prediction was the charge conjugation symmetry.\footnote{A notion that soon afterwards was generalised to all spins.} In 1937 Majorana introduced a new class of particles for which the particles and antiparticles were identical~\cite{Majorana:1937vz}. They were described by a field which equaled its charge conjugate. The new field was still based on Dirac spinors, which we shall see below are eigenspinors of the parity operator.

Fast forward a few decades, with the intervening years placing Dirac's work on a more systematic footing,
the new astrophysical and cosmological observations have now introduced a new duplexity. These observations strongly hint that there exists a new form of matter which carries no, or limited, interactions with the matter and gauge fields of the standard model of high energy physics~\cite{Bertone:2016nfn}. The new form of matter, to distinguish it from the matter fields of the standard model of high energy physics, is called dark matter. It  poses a new duplexity.
For some decades now supersymmetry was thought to provide precisely such a duplexity phenomena in a natural manner by introducing a symmetry that transmuted mass dimensionality and statistics of particles. However, despite intense searches, there is no observational evidence for its existence.

At present there is no clear cut understanding of what populates the dark sector. Its existence is known only from its gravitational imprints and its implications for the cosmic structure formation~\cite{Blumenthal:1984bp}. Taking a hint from the standard model we here explore the conjecture that the matter fields for dark matter are fermionic. 

Whatever these fermions are they must still be one representation or the other of the spacetime symmetries~\cite{Wigner:1939cj}. If one allows for the possible existence of  a new symmetry that mutates the mass dimensionality of fermions, without affecting the statistics, then the no go theorems resulting from the works of Wigner, Weinberg, Lee and 
Wick~\cite{Wigner:1962ep,PhysRev.133.B1318,Weinberg:1995mt,PhysRev.148.1385} no longer apply.
This is what we do here and construct a quantum theory of mass dimension one fermions. 

In a parallel with Majorana fermions, these new fermions do not allow the usual local gauge symmetries of the standard model. Concurrently, their mass dimension is in mismatch with mass dimension three halves of the standard model fermions. It  prevents them from entering the standard model doublets. Combined, these aspects make the new fermions a first principle dark matter candidate. 

Like the Majorana fermions the symmetry of charge conjugation plays a central role for the new fermions: while for the Majorana field the coefficient functions are eigenspinors of the parity operator the field itself equals its charge conjugate,  for the new fermions the field is expanded in terms of the eigenspinors of the charge conjugation operator. 
Once that is done, one may choose to impose the Majorana condition, but it is not mandated.

Thus the new formalism, in a parallel with the Dirac formalism, allows for darkly-charged fermions, and Majorana-like neutral fields.

The doubling of the degrees of freedom for the  spin half fermions of Dirac can be traced to its parity covariance. This symmetry requires not only the left-handed Weyl spinors but also right-handed Weyl spinors. In the process for a spin one half particle we are forced to deal with four, rather than two, degrees of freedom. The antiparticles of the Dirac formalism are an astounding consequence of this doubling. 

As noted above, Dirac and Majorana quantum field are both expanded in terms of Dirac spinors. These are eigenspinors of the parity operator: $m^{-1} \gamma_\mu p^\mu$~\cite{Speranca:2013hqa}. Eigenspinors of the charge conjugation operator are thought to provide no Lagrangian description in a quantum 
field theoretic construct~\cite[App. P]{Aitchison:2004cs}. Thus placing the parity and charge conjugation symmetries on an asymmetric footing -- that is, as far as their role in constructing spin one half quantum fields are concerned. Here we show that this claim is in error. It has remained hidden in a lack of full appreciation as to how one is to construct duals for spinors, and the associated adjoints -- that is, in the mathematics underling the definition of the spinors via
$\overline\psi(\p) = \psi(\p)^\dagger \gamma_0$. 

To develop the physics hinted above we are forced to complete the development of this mathematics. Taken to its logical conclusion
it leads to the doubling of the fundamental form of matter fields that  is supported by Lorentz symmetries. One form of matter is described by the Dirac formalism, while the other, that of the dark sector, by the new fermions reported here. For each sector the needed matter fields 
 require a complete set of four, four-component spinors. For the former these are eigenspinors of the  parity operator while for the latter these are eigenspinors of the  charge conjugation operator. Global phases associated with these eigenspinors, and the pairing of these eigenspinors with the creation and annihilation operators, influence the Lorentz covariance and locality of the fields. This last observation, often ignored in textbooks, when coupled with the 
discovery of a freedom in defining spinorial duals 
 accounts for removal of the non-locality and restoring of the Lorentz symmetry for the mass dimension one fermions of~\cite{Ahluwalia:2004sz,Ahluwalia:2004ab}.

The new fermions are totally unexpected. 
Like supersymmetry  the new formalism transmutes the mass dimensionality of spin half particles to one, but without altering the statistics.

This communication supersedes all our previous publications on the subject and for the first time provides a Lorentz covariant local quantum field of mass dimension one fermions of spin one half.

\subsection{On the presentation of the paper\label{sec-1.1}}

The chosen title, and its resemblance with the titles in references~\cite{Dirac:1928hu,PhysRev.76.749}, reflects the fact that we are introducing an hitherto unexpected new class of particles: fermions with mass dimension one.

Following this section, section~\ref{sec:notation} sets up the notation.

Without reference to a wave equation or a Lagrangian density,
in section~\ref{sec:parity} we present a discussion of the parity operator for the four-component 
spinors, while in section~\ref{sec:Dirac-spinors} we
show that the Dirac spinors are eigenspinors of the parity operator. In section~\ref{sec:Dirac-spinors} 
we also share with the reader insights from Weinberg's theory of quantum fields (ToQFs, ~\cite{Weinberg:1995mt}). These are extremely important for the work that we undertake here.

In section~\ref{sec:charge-conjugation-operator} we construct the charge conjugation operator for the four-dimensional representation space of the four-component spinors. Again, 
without reference to a wave equation or a Lagrangian density.
Then,  in
section~\ref{sec:elko}  we introduce a complete set of c-number eigenspinors of the charge conjugation operator, Elko (a German acronym for 
{\textbf{E}}igenspinoren des {\textbf{L}}adungs{\textbf{ko}}njug\-ationsoperators 
introduced in~\cite{Ahluwalia:2004sz,Ahluwalia:2004ab}).
It ends
with a brief remark on global phase transformation of the eigenspinors of the charge conjugation operator. In section~\ref{sec:elko-explicit} we construct the new spinors explicitly. 
It becomes apparent there that the boost does not mix  various spinorial components -- instead, it simply scales them through two energy dependent factors. This is in sharp contrast to the Dirac case. This section also makes explicit various phases that prove to be important for the locality properties of the new quantum field. 
 The hint for the new mass dimension one quantum fields first emerges in section~\ref{sec:elko-do-not-satisfy-Dirac-equation} where we show that the new spinors do not satisfy Dirac equation. Section \ref{sec:parenthetic-observation} is devoted to discrete symmetries for Elko, including the time reversal operator. 
 
Having developed the Elko formalism, in section~\ref{sec:the-mathematical-theory-of-duals} we motivate the need for developing a mathematical theory of spinorial duals in a step by step process through sections~\ref{sec:temptation}, \ref{sec:The-questions-on-the-path-of-departure}, and \ref{sec:A-need-for-a-new-dual-for-Elko}.
Section~\ref{sec:constraint-from-scalar-invariants} obtains the constraint on duals from the invariance of the inner product and applies it to Dirac spinors in section~\ref{sec:Dirac-dual}, and in section~\ref{sec:elko-dual} to Elko.
Section~\ref{sec:equivalent-representations} provides the equivalence of the results thus obtained with earlier, but less systematic, considerations. In section~\ref{sec:constraints-from-the-spin-sums} we derive constraints on the Elko dual from the invariance of the Elko spin sums.

The new quantum field with Elko as its expansion coefficients, its adjoint, Feynman-Dyson propagator, Lagrangian density, and  zero point energy are introduced in section~\ref{sec:Elko-quantum-field}. Interactions of the new fermions are discussed in section~\ref{sec:interactions}.
Section~\ref{sec:locality} establishes the locality of the new mass dimension one field. This is followed by section~\ref{sec:Majorana-isation} where we implement ``Majorana-isation'' of the new field. 

A guide to the existing literature on Elko and mass dimension one fermions appears in section~\ref{sec:literature}.

The conclusion in section~\ref{sec:conclusion} takes the form of  a brief summary   of the task accomplished 
by this communication.\footnote{ While much of the work presented here is new we have not hesitated in freely borrowing results from our previous calculations when that adds to a self-contained smooth flow of the narrative. The material presented here touches on the foundations of physics and for that reason we have made an effort to keep the tone and content of our presentation on the pedagogic side.  }

\subsection{Setting the notation: Weyl spinors\label{sec:notation}}

The right- and left-handed Weyl spinors transform under Lorentz boost as
[see remarks after equation~(\ref{eq:jxjyjz})]
\begin{align}
		\phi_R(p^\mu) = \exp\left(+\frac{\s}{2}\cdot\vp\right) 
		\phi_R(k^\mu),\quad
		\phi_L(p^\mu) = \exp\left(-\frac{\s}{2}\cdot\vp\right) \phi_L(k^\mu)\label{eq:w}
\end{align}
where $\s$ represents the set  of Pauli matrices, $(\sigma_x,\sigma_y,\sigma_z)$ in their standard representation and the boost parameter $\vp$ is defined so that $\exp\left(i \K\cdot\vp\right)$ acting on the standard four momentum
\begin{equation}
		k^\mu \stackrel{\rm def}{=} \left(m,\lim_{p\rightarrow 0}\frac{\p}{p}\right),\quad p = \vert\p\vert \label{eq:def-k}
\end{equation} 
equals the general four momentum 
$
			p^\mu = (E, p \sin\theta\cos\phi,p\sin\theta\sin\phi,p\cos\theta).
$
This yields
$\cosh\varphi = E/m$, $\sinh\varphi= p/m$ with $\widehat{\vp} = \widehat {\p}$, while
$\K$ are the $4\times 4$ matrices  for the generators of boosts in Minkowski space:
\footnote{We shall use the conventions of~\cite{Ryder:1985wq} with $\hbar$ and $c$ set to unity. } 

\begin{align}
K_x &=\left[
\begin{array}{cccc}
0 & -i & 0 & 0\\
-i & 0 & 0 & 0\\
0 &0 & 0 & 0\\
0 &0 & 0 & 0\\
\end{array}
\right], \quad K_y=\left[
\begin{array}{cccc}
0 & 0 & -i  & 0\\
0 & 0 & 0 & 0\\
-i &0 & 0 & 0\\
0 &0 & 0 & 0\\
\end{array}
\right], \nonumber\\
K_z &=\left[
\begin{array}{cccc}
0 & 0 & 0  & -i \\
0 & 0 & 0 & 0\\
0 &0 & 0 & 0\\
-i &0 & 0 & 0\\
\end{array}
\right].  \label{eq:Minkowski-boost-generator}
\end{align}
The three generators of rotation are
\begin{align}
J_x &=\left[
\begin{array}{cccc}
0 & 0 & 0 & 0\\
0 & 0 & 0 & 0\\
0 &0 & 0 & -i\\
0 &0 & i & 0\\
\end{array}
\right], \quad J_y=\left[
\begin{array}{cccc}
0 & 0 & 0  & 0\\
0 & 0 & 0 & i\\
0 &0 & 0 & 0\\
0 &-i & 0 & 0\\
\end{array}
\right], \nonumber\\
 J_z &=\left[
\begin{array}{cccc}
0 & 0 & 0  & 0 \\
0 & 0 & -i & 0\\
0 &i & 0 & 0\\
0&0 & 0 & 0\\
\end{array}
\right].  \label{eq:jxjyjz}
\end{align}

Equations (\ref{eq:w})  follow from the fact that $- i \s/2$ are the generators of the boosts for the right-handed Weyl representation space, while $+ i \s/2$ are  for the left-handed Weyl representation space. 
For the direct sum of the right- and left-Weyl representation spaces, to be motivated below,
the boost and rotation generators thus read
\begin{equation}
				\kb  = \left[
				\begin{array}{cc}
				- i \s/2 & \0\\
					\0 & + i \s/2
					\end{array}
					\right], \quad
 \bz = \left[
\begin{array}{cc}
 \s/2 & \0\\
\0 &  \s/2
\end{array}
\right].  \label{eq:pi}
\end{equation}

The set of generators $\{\K,\J\}$ and $\{\kb,\bz\}$, separately, satisfy the same unifying algebra, the Lorentz algebra 
\begin{align}
&\left[\mathfrak{J}_x,\mathfrak{J}_y \right]= i \mathfrak{J}_z,\quad\mbox{and cyclic permutations}\nonumber\\
&\left[\mathfrak{K}_x,\mathfrak{K}_y\right] = - i\mathfrak{J}_z,\quad\mbox{and cyclic permutations}\nonumber\\
&\left[\mathfrak{J}_x,\mathfrak{K}_x \right]=0,  \quad\mbox{etc.}\nonumber\\
&\left[\mathfrak{J}_x,\mathfrak{K}_y \right]= i \mathfrak{K}_z,\quad\mbox{and cyclic permutations}\nonumber\\
&\left[\mathfrak{K}_x,\mathfrak{J}_y\right] =  i\mathfrak{K}_z,\quad\mbox{and cyclic permutations}
\end{align}
and are simply its different representations. The ${\mathfrak{J}}_i$ and $\mathfrak{K}_j$, $i,j=x,y,z$, represent generators of rotations and boosts. Their exponentiations
\begin{equation}
\exp\left(i\bmfk\cdot\vp\right), \quad \exp\left(i\bmfj\cdot\bm{\theta}\right)
\end{equation}
give the transformations under  boosts and rotations for the `vectors' spanning the associated representation space.

The ${\Lambda^\mu}_\nu$ representing boosts and rotations in Minkowski space is thus given by
\begin{equation}
\Lambda \stackrel{\mathrm{def.}}{=}\left\{
\begin{array}{cl}
\exp\left(i\K\cdot\vp\right) & \mbox{for Lorentz boosts} \\
\exp\left(i\J\cdot\vp\right) & \mbox{for rotations}
\end{array}\right.\label{eq:Minkowski-boost-rotation}
\end{equation}
The elements of $\Lambda$, $\K$, and $\J$ are of the form ${{a}^\mu}_\nu$, where the 
$4\times 4$ matrix $a$ stands generically for either one of them.

\section{Parity operator for  general four-component spinors
\label{sec:parity}}

In the Minkowski space, Parity is defined as the map
\begin{equation}
P: x^\mu = (x^0, \x) \rightarrow x^{\prime\mu} =(x^0, -\x).\label{eq:parity-minkowski}
\end{equation}
An \textit{ab initio} examination of the question, ``How does it affect the spinor spaces?'' yields the Dirac operator~\cite{Speranca:2013hqa}.  The argument is as follows.

 A  general  4-component spinor is a direct sum of the $(1/2,0)$ and $(0,1/2)$ Weyl spinors (mass $m\neq 0$, throughout this communication)
 \begin{equation}
		\psi(p^\mu) =\left( \begin{array}{c}
		\phi_R(p^\mu)\\
		\phi_L(p^\mu)
		\end{array}
		\right) = \exp(i \kb\cdot\vp) \psi(k^\mu)\label{eq:psi}
\end{equation}
with $p^\mu = \exp\left(i\K\cdot\vp\right)k^\mu$. With $\K$ given by 
equations~(\ref{eq:Minkowski-boost-generator}),  $\exp\left(i\K\cdot\vp\right)$ is simply the usual 
${\Lambda^\mu}_\nu$.
These spinor may be eigenspinors of the parity operator, or that of the  charge conjugation operator, or of any other physically or mathematically relevant operator -- which, incidentally, may make the 
right- and left- transforming components in $\psi(p^\mu)$ loose their independence. A physically important classification of spinors is by Lounesto~\cite[Chapter 12]{Lounesto:2001zz}.

The $\psi(k^\mu)$ are generally called  ``rest  spinors,'' while the $\psi(p^\mu)$ are often named ``boosted spinors.'' However, since no frame is a preferred frame, the $\psi(k^\mu)$ and  the infinitely many $\psi(p^\mu)$ reside in every frame. 

For the sake of completeness it is worth noting that the rotation on $\psi(p^\mu)$  is implemented by
\begin{equation}
\psi\left(p^{\prime\mu}\right) = \exp\left(i\bz\cdot\bm{\theta}\right) \psi\left(p^{\mu}\right)\label{eq:rotation-on-spinors}
\end{equation}
with $\theta$ as the angle which characterises  the rotation, about an axis $\widehat n$
\begin{equation} \bm{\theta} = \theta\, \widehat n
\end{equation} and $p^{\prime\mu} = \exp\left(i\J\cdot\bm{\theta}\right) p^\mu$.

Under  $P$, $\s\to \s$ and $\vp\to - \vp$, therefore
 $P$ interchanges the right- and left- handed Weyl representation spaces 
 \begin{equation}
 (1/2,0) \stackrel{P}{\longleftrightarrow} (0,1/2).
 \end{equation}
 The effect of $P$ on $\psi(p^\mu)$ is thus realised by a $4\times 4$ matrix $\mathcal{P}$:
 
 \begin{itemize} 
 \item[a.]
which
 up to a global phase
must contain purely off-diagonal $2\times 2$ identity matrices $\openone$, and
\item[b.]
which in addition implements the action of $P$ on $p^\mu$.  
\end{itemize}
Up to a global phase, chosen to be $1$ (in much of this communication), the effect of $\mathcal{P}$ on $\psi(p^\mu)$ is therefore
given by
\begin{equation}
		\mathcal{P}\, \psi(p^\mu) =
		\left(\begin{array}{cc}
		\0 & \I\\
		\I & \0
		\end{array}
		\right)\psi(p^{\prime\mu}) = \gamma_0 \psi(p^{\prime\mu}). \label{eq:psi2}
\end{equation}
Here, $p^{\prime\mu}$ is the $P$ transformed $p^\mu$ while $\0$ and $\I$ represent $2\times 2$ null and identity matrices, respectively. This is where the general textbook considerations on $\mathcal{P}$ stop.

 For a general spinor, Speran\c{c}a has noted that 
it provides a better understanding of $\mathcal{P}$ if in (\ref{eq:psi2}) we note that
 $ \psi(p^{\prime\mu}) $ may be related to  $ \psi(p^\mu)$ as follows~\cite{Speranca:2013hqa}
\begin{equation}
 	\psi\left(p^{\prime\mu}\right) = \exp\big[i \kb\cdot(-\vp)\big]\psi(k^\mu) 
\end{equation}
with  $\kb$ defined in (\ref{eq:pi}). But since from (\ref{eq:psi}), $\psi(k^\mu)  = \exp\left(-i\kb\cdot\vp\right)\psi(p^\mu)$ the above equation can be re-written as
\begin{equation}
\psi\left(p^{\prime\mu}\right)\,		 =  
\exp\left(-i \kb\cdot\vp\right)\exp\left(-i \kb\cdot\vp\right) \psi(p^\mu)
		  =  \exp(- 2 i \kb\cdot\vp) \psi(p^\mu). \label{eq:zimpok167} 
\end{equation} 
Substituting $\psi\left(p^{\prime\mu}\right)$ from the above equation in (\ref{eq:psi2}), and on using the anti-commutativity of $\gamma_0$ with each of the generators of the boost, 
\begin{equation}
\{\gamma_0,\kb_i\} = 0,\qquad  \mbox{with}\,\, i=x,y,z
\end{equation} 
equation (\ref{eq:psi2}) becomes
\begin{equation}
		\mathcal{P} \psi(p^\mu) = \exp(2 i \kb\cdot\vp) \gamma_0 \psi(p^\mu).\label{eq:psi3}
\end{equation}

A direct evaluation of the exponential in 
the right hand side of the above equation gives the identity
 \begin{equation}
 		\exp\left( 2 i \kb\cdot\vp\right) \gamma_0 = m^{-1} \gamma_\mu p^\mu \label{eq:zimpok177}
 \end{equation}
where $\gamma_\mu=(\gamma_0,\bm{\gamma})$ are the Dirac matrices in the Weyl representation
\begin{equation}
\gamma_0 = \left(\begin{array}{cc}
\0 & \openone \\
\openone & \0\end{array}\right),\quad
 \bm{\gamma} = \left(\begin{array}{cc}
 \0 &\s \\
 -\s & \0
 \end{array}\right).\label{eq:diracgamma-lower}
 \end{equation}
  A substitution of expansion (\ref{eq:zimpok177}) in (\ref{eq:psi3}) results in
 \begin{equation}
		\mathcal{P} \, \psi(p^\mu) = m^{-1}\gamma_\mu p^\mu \, \psi(p^\mu).\label{eq:psi4}
\end{equation}
Up to the above-mentioned  global phase this exercise yields the parity operator
\begin{equation}
		\mathcal{P} = m^{-1} \gamma_\mu p^\mu. \label{eq:P}
\end{equation}

\subsection{Dirac spinors, their phases, and their pairing with the creation and annihilation operators \label{sec:Dirac-spinors}}

The $\mathcal{P}$ applies to all spinors, but its eigenspinors are what we generally call as Dirac spinors. The eigenvalues of $\mathcal{P}$ are
 $\pm 1$. Each of these has a  two fold degeneracy 
\begin{equation}
		\mathcal{P}\, \psi^S_\sigma(p^\mu) = + \psi^S_\sigma(p^\mu),\quad
		\mathcal{P}\, \psi^A_\sigma(p^\mu) = - \psi^A_\sigma(p^\mu). \label{eq:Dirac-p}
\end{equation}
The subscript $\sigma$ is the degeneracy index, while the 
superscripts refer to self and anti-self conjugacy of $\psi(p^\mu)$ under $\mathcal{P}$. This nomenclature is helpful for pedagogic reasons.  With the help of Eq.~(\ref{eq:P}), Eq.~(\ref{eq:Dirac-p}) translates to
\begin{equation}
		\left(\gamma_\mu p^\mu -m \openone\right) \psi^S_\sigma(p^\mu) = 0,\quad
		\left(\gamma_\mu p^\mu +m\openone\right) \psi^A_\sigma(p^\mu) = 0.
\end{equation}

The Dirac's~\cite{Dirac:1928hu} $u_\sigma(p^\mu)$ and $v_\sigma(p^\mu)$ spinors are thus seen as the eigenspinors of the parity operator, $\mathcal{P}$, with eigenvalues $+1$ and $-1$, respectively:
\begin{equation}
	\psi^S_\sigma(p^\mu) \to u_\sigma(p^\mu),\quad  \psi^A_\sigma(p^\mu)\to v_\sigma(p^\mu).
\end{equation}
We end this part of the discussion on the spinorial parity operator by noting that $
\mathcal{P}^2 = \I _4$, 
and that the form of equation (\ref{eq:Dirac-p}) is preserved under a global transformation
\begin{equation}
			\psi_\sigma(p^\mu) \to  \psi_\sigma^\prime(p^\mu)= \exp( i \alpha) \psi_\sigma(p^\mu),\quad
			 \forall \sigma
			\label{eq:remarks}
\end{equation}
with $\alpha\in\mathfrak{R}$. This primitive fact, with $\alpha$ independent of spacetime and of the degeneracy index $\sigma$,  allows the introduction of a \textit{local} $U(1)$ gauge interaction when the eigenspinors are of $\mathcal{P}$ are used as expansion coefficients of a fermionic spin one half field, that is the Dirac field.

Lest one be misled, it must be noted that while constructing a quantum field with the Dirac spinors as its expansion coefficients the phases that the spinors can carry are completely fixed by the combined demands of Lorentz and parity covariance, and that of locality~\cite{Weinberg:1995mt}. Strictly speaking, this statement is true up to a freedom of a global phase for the field. The said constraints also fix the pairing of the Dirac spinors with the creation and annihilation operators. Many respected textbooks on quantum field theory, surprisingly, violate the indicated pairing as well as make a (wrong)choice for the spinor phases that on Majorana-isation of the field reveal a serious internal inconsistency (one of these being, non-locality). 
After the appearance of the Steven Weinberg's classic on ToQFs  these foundational flaws are beginning to evaporate from some of the textbooks that have followed.

The above remarks refer as much as to other, otherwise distinguished, authors as to us. Much of the non-locality in our 2005 work~\cite{Ahluwalia:2004sz,Ahluwalia:2004ab} had to do with our own lack of appreciation of the just stated remarks (see section~\ref{sec:elko-explicit} below).

A heuristic way to motivate that the Dirac spinors, when used as expansion coefficients for a quantum field, must loose the freedom of multiplicative global phases is as follows. Each of the four spinors in the set 
$\{u_\sigma(p^\mu),v_\sigma(p^\mu)\}$ remains a solution of the Dirac equation if it is multiplied by a 
$\sigma$-dependent global phase
\begin{align}
&u_\sigma(p^\mu) \to \exp( i\alpha_\sigma) u_\sigma(p^\mu),\quad
v_\sigma(p^\mu) \to \exp( i\beta_\sigma) v_\sigma(p^\mu)\nonumber\\
&\alpha,\beta\in \Re.
\end{align}
However, as soon as these spinors are used as expansion coefficients for a field, the global phases cease to be so for a field (unless one had chosen all four of them to be same). These phases must be chosen, along with their pairing with the creation and annihilation operators, to yield a Lorentz and parity covariant local field. This is a rough insight that one gains on reading the first few chapters of Weinberg's ToQFs. Majorana-isation of the Dirac field introduces additional `relativisation of the global phases,' and these new phases make the resulting field non-local. As an example, all one has to do is  to pick up the Dirac field as found in reference~\cite{Ryder:1985wq} and set $d_\sigma(p^\mu)^\dagger
= b_\sigma(p^\mu)^\dagger$, and evaluate the three locality anti-commutators. One would then immediately find the said non-locality.

The heuristic exercise just outlined above was first undertaken by Tom Watson, then an undergraduate student at the University of Canterbury,  at the suggestion of the author. It  was soon afterwards confirmed by Cheng-Yang Lee in the same research group~\cite{Cheng-Yang-Lee:2007}.

 \section{Charge conjugation operator for four-component spinors\label{sec:charge-conjugation-operator}}

As already mentioned in the opening section, introduction of the four-compo\-nent spinors (\ref{eq:psi}), not necessarily as eigenspinors of the parity operator, doubles the degrees of freedom from two, for spin one half, to four. In a quantum field theoretic formalism this doubling introduces the notion of antiparticles and is required for preserving causality~(see,\cite[Section 2.13]{Weinberg:1972kfs} and \cite{Feynman:1987gs}).
 The  charges of the particles and antiparticles are then determined by the type of local gauge symmetries that the underlying kinematic framework supports. 
 
  The particle-antiparticle symmetry enters via  charge conjugation operator.
For the four-component spinors, it may be similarly constructed as $\mathcal{P}$ without first invoking a wave equation or a Lagrangian density. To see this we begin with the observation that the Wigner time reversal operator for spin one-half, $\Theta$,
acts on the Pauli matrices as follows\footnote{\label{fn:Theta}For any spin, $\Theta_{[j]} \J \Theta_{[j]}^{-1} = - \J^\ast$; with $\Theta_{[j]} = (-1)^{j+\sigma}\delta_{\sigma^\prime,-\sigma}$ and $\Theta_{[j]}^\ast \Theta_{[j]} = (-1)^{2 j}$. For convenience, we abbreviate $\Theta_{\left[1/2\right]}$ to $\Theta$. In $\Theta$ we mark the rows and columns in the order $\{-1/2,1/2\}$.}
	\begin{equation}
		\Theta \s\Theta^{-1} = - \s^\ast\label{eq:wt}
	\end{equation}
with 
	\begin{equation}
			\Theta= \left(\begin{array}{cc} 0 & -1 \\ 1 & 0\end{array}\right), \quad
			\Theta^{-1} = - \Theta.
	\end{equation}
As we show below, it allows the following `magic'  to happen: 
\begin{itemize}
\item[1.]
If $\phi_L(p^\mu)$ transforms as a left-handed Weyl spinor in accordance with (\ref{eq:w}) then $\zeta_\lambda \Theta \phi^\ast_L(p^\mu)$ transforms as a right-handed Weyl spinor, where $\zeta_\lambda$ is an arbitrary phase,
and 
\item[2.]
Similarly, if $\phi_R(p^\mu)$ transforms as a right-handed Weyl spinor then $\zeta_\rho \Theta \phi^\ast_R(p^\mu)$ transforms as a left-handed Weyl spinor, where $\zeta_\rho$ is another arbitrary  phase.
\end{itemize}

This important result arises as follows:
First complex conjugate both the equations in (\ref{eq:w}), then multiply from the left by $\Theta$, and use the above defining feature of the Wigner time reversal operator. This sequence of manipulations (after using the freedom to  multiply these equations by phases $\zeta_\lambda$  and $\zeta_\rho$ respectively) ends up with the result
\begin{align}		
	  &\Big[\zeta_\lambda\Theta \phi^\ast_L(p^\mu)\Big] = \exp\left( + \frac{\s}{2}\cdot\vp\right)\Big[\zeta_\lambda\Theta \phi^\ast_L(k^\mu)\Big]\nonumber\\
	  &\Big[\zeta_\rho\Theta \phi^\ast_R(p^\mu)\Big] = \exp\left( - \frac{\s}{2}\cdot\vp\right)\Big[\zeta_\rho\Theta \phi^\ast_R(k^\mu)\Big]
\end{align}
 and yields the claimed magic of the Wigner time reversal operator. This crucial  observation motivates  the  introduction of  two sets of four-component spinors~\cite{Ahluwalia:1994uy}
	\begin{equation}
	\lambda(p^\mu) = \left[\begin{array}{c}
					\zeta_\lambda \Theta \phi^\ast_L(p^\mu)\\
					\phi_L(p^\mu)
					\end{array}
					\right],\quad
					\rho(p^\mu) = \left[\begin{array}{c}
									\phi_R(p^\mu)\\
					\zeta_\rho \Theta \phi^\ast_R(p^\mu)
					\end{array}
					\right].\label{eq:lambda}
	\end{equation}
The $\rho(p^\mu)$ do not provide an additional independent set of  spinors  and for that reason  we do not consider them further and confine our attention to $\lambda(p^\mu)$ only~\cite{Ahluwalia:2004ab}. 

Generally,  part of this result is introduced as a `magic of Pauli matrices'~\cite{Ramond:1981pw} where $ \Theta$ gets concealed in Pauli's $\sigma_y$,  which  equals $i \Theta$. Our argument  in terms of the Wigner time reversal operator $\Theta$ has the advantage that it immediately generalises  to higher spins. Furthermore, the recognition that there is an element of freedom in the indicated phases, $\zeta_\lambda$ and $\zeta_\rho$, makes $\lambda(p^\mu)$ escape their interpretation as 
Weyl spinors in a four-component disguise. This happens because one can now have four rather than two independent four-component spinors (see below).

With these observations at hand we are  led to entertain the possibility that in addition to the symmetry operator $\mathcal{P}$, there may exist a second symmetry operator, which up to a global phase factor, has the form
	\begin{equation}
	\mathcal{C} \stackrel{\mathrm{def}}{=} \left[\begin{array}{cc}
					\0 & \alpha \Theta \\
					\beta\Theta & \0
					\end{array}
				\right] K \label{eq:cc}
	\end{equation}
where $K$ complex conjugates to its right. The arguments that leads to~(\ref{eq:cc}) are similar to the ones that give~(\ref{eq:psi2}). Requiring $\mathcal{C}^2$ to be an identity operator determines $\alpha = i, \beta= -i$ (where we have used $K^2=1$). It results in
	\begin{equation}
	\mathcal{C} = \left[\begin{array}{cc}
					\0 & i\Theta \\
					-i \Theta & \0
					\end{array}
					\right] K = \gamma_2 K.\label{eq:gamma2K}
	\end{equation}
There also exists a second solution with $\alpha = - i, \beta= i$. But this does not result in a physically different operator and in any case the additional minus sign can be absorbed in the indicated global phase. This is the same operator that appears in the particle-antiparticle symmetry associated with the 1928 Dirac equation~\cite{Dirac:1928hu}. 

We have thus arrived at the charge conjugation operator from the analysis of the symmetries of the $4$-component representation space of spinors. This perspective has the advantage of immediate generalisation to any spin: if $\Theta_{[j]}$ is taken as Wigner time reversal operator for spin $j$ then the spin-$j$ charge conjugation operator in the $2(2j+1)$ dimensional representation space becomes~\cite[equation A10]{Lee:2012td}
	\begin{equation}
		\mathcal{C} = \left[\begin{array}{cc}
		\0 & -i \Theta_{[j]}^{-1} \\
		-i \Theta_{[j]} & \0
		\end{array}
		\right] K. \label{eq:c}
	\end{equation}
For spin one half, $\Theta^{-1} = - \Theta$; consequently, the above expression 
coincides with the result for spin one half given in equation~(\ref{eq:gamma2K}).

Both $\mathcal{P}$ and $\mathcal{C}$ arise without any reference to one wave equation or the other, or equivalently without assuming one Lagrangian density or another. In fact, as will become clear from our presentation, Lagrangian densities are something one must derive rather than assume.

\subsection{Eigenspinors of the charge conjugation operator (Elko)\label{sec:elko}}

The presence of $K$ in the definition of the charge conjugation operator~(\ref{eq:gamma2K})
introduces an element of freedom in choosing the eigenvalues of $\mathcal{C}$ 
	\begin{align}
		\mathcal{C} \lambda(p^\mu) & = 
		\left(\begin{array}{cc}
					\0 & i\Theta \\
					-i \Theta & \0
					\end{array}
					\right) 
					\left(\begin{array}{c}
					\zeta^\ast_\lambda \Theta \phi_L(p^\mu)\\
					\phi^\ast_L(p^\mu)
					\end{array}
					\right) \nonumber\\
					&=  \left(\begin{array}{cc}
					 i \Theta \phi^\ast_L(p^\mu)\\
					 - i \zeta^\ast_\lambda \Theta^2 \phi_L(p^\mu)
			\end{array}	\right)	 \nonumber\\
					&
			 =
					 \left(\begin{array}{cc}
					 i \Theta \phi^\ast_L(p^\mu)\\
					 i \zeta^\ast_\lambda \phi_L(p^\mu)
			\end{array}			 \right).	
		\label{eq:Cev}		
	\end{align}
The choice $\zeta_\lambda = \pm i$ makes $\lambda(p^\mu)$ become eigenspinors of $\mathcal{C}$ 
with doubly degenerate eigenvalues $\pm 1$:
\begin{equation}
	\mathcal{C}  \lambda^S(p^\mu) = +  \lambda^S(p^\mu),\quad
	\mathcal{C}  \lambda^A(p^\mu) = -  \lambda^A(p^\mu),    \label{eq:elko-cc}
	\end{equation} 
where	 
	 	\begin{equation}
  		\lambda(p^\mu) = \left\{
		\begin{array}{ll}
	 	\lambda^S(p^\mu) =
	\left(\begin{array}{c}  i \Theta \phi_L^\ast(p^\mu)\\
	                       \phi_L(p^\mu)
	                       \end{array} 	\right)	
		 & \mbox{for} ~ \zeta_\lambda = + i \\ \\  
	 	
		\lambda^A(p^\mu) =
		\left(\begin{array}{c}-  i \Theta \phi_L^\ast(p^\mu)\\
	                       \phi_L(p^\mu)
	                       \end{array} 	\right)	 &  \mbox{for} ~ \zeta_\lambda = - i   
		\end{array}\right.
		\label{eq:lsa}
 	 \end{equation}

Because of the presence of the operator $K$ in Eq.~(\ref{eq:gamma2K})
a global transformation of the type
\begin{equation}
			\lambda(p^\mu) \to  \lambda^\prime(p^\mu)= \exp( i \mathfrak{a} \alpha) 
			\lambda(p^\mu) \label{eq:counterpart}
\end{equation}
with $\mathfrak{a}^\dagger = \mathfrak{a}$, a $4\times 4$ matrix and $\alpha\in\mathfrak{R}$, does not preserve the self/anti-self conjugacy of $\lambda(p^\mu)$ under $\mathcal{C}$ unless
the matrix $\mathfrak{a}$ satisfies the condition
\begin{equation}
		\gamma_2 \mathfrak{a}^\ast +\mathfrak{a} \gamma_2 = 0
\end{equation}
The general form of $\mathfrak{a}$ satisfying these requirements is found to be
\begin{align}
		\mathfrak{a} & = 
		\left[
		\begin{array}{cccc}
		 \epsilon  & \beta  & \lambda & 0 \\
		 \beta^\ast  & \delta  & 0 & \lambda  \\
 	\lambda^\ast  & 0 & -\delta  & \beta  \\
 	0 & \lambda^\ast  & \beta^\ast  & -\epsilon  \\
		\end{array}
		\right] 
		\label{eq:mathfraka}
\end{align}
with $ \epsilon,\delta \in \mathfrak{R}$ and $\beta,\lambda \in\mathfrak{C}$ (with no association with the same symbols used elsewhere in this work). 

For a field constructed with the eigenspinors of $\mathcal{C}$ as expansion coefficients,
the usual local $U(1)$ interaction is ruled out as 
 the form of $\mathfrak{a}$ given by (\ref{eq:mathfraka}) does not allow a solution with $\mathfrak{a}$ proportional to an identity matrix.
As remarked around equation (\ref{eq:remarks}), for the eigenspinors of the $\mathcal{P}$, defined by (\ref{eq:Dirac-p}), the counterpart of (\ref{eq:counterpart} ) is trivially satisfied. And thus the two fields, one based on $\mathcal{P}$ eigenspinors and the other constructed from the $\mathcal{C}$ eigenspinors, carry intrinsically different possibility for their interaction through local gauge fields. The simplest non-trivial choice consistent with (\ref{eq:mathfraka}) is given by 
\begin{equation}
\mathfrak{a} = \gamma = \frac{i}{4!} \epsilon_{\mu\nu\lambda\sigma}
\gamma^\mu\gamma^\nu\gamma^\lambda\gamma^\sigma = \left(\begin{array}{cc}
\I &\0\\
\0 & -\I
\end{array}\right)
\end{equation}
where $\epsilon_{\mu\nu\lambda\sigma}$ is defined as
\begin{equation}
\epsilon_{\mu\nu\lambda\sigma} = \left\{
\begin{array}{cl} 
+1, &  \mbox{for  $\mu\nu\lambda\sigma$ even permutation of 0123}\\
- 1,&  \mbox{for  $\mu\nu\lambda\sigma$ odd permutation of 0123}\\\
0, & \mbox{if any two of the $\mu\nu\lambda\sigma$ are same}
\end{array}\right.
\end{equation}

\subsection{Explicit construction of $\bm{{\lambda(p^\mu)}}$ and locality 
phases\label{sec:elko-explicit}}

To obtain an explicit form of $\lambda(p^\mu)$ calls for a choice of the `rest' spinors, $\lambda(k^\mu)$, with $k^\mu$ defined in (\ref{eq:def-k}). That done, one then has for an arbitrary $p^\mu$
\begin{equation}
	\lambda(p^\mu) = \exp(i \kb\cdot\vp) \lambda(k^\mu).   \label{eq:elkoboost}
\end{equation}
In principle, as remarked above, the boosted spinors reside in the boosted frames. But since no frame is a preferred frame they must also exist in all frames (an argument generally attributed  to E. P. Wigner\footnote{The author would appreciate if the readers may provide a specific reference in published literature.}). It is this interpretation that we attach to $\lambda(p^\mu)$.
With the generator of the boost, $\kb$, defined in (\ref{eq:pi}), the boost operator in 
(\ref{eq:elkoboost}) can be readily evaluated using $(\s\cdot\widehat\p)^2 = \openone$, to the effect that
\begin{align}
 \exp(i \kb\cdot\vp)  & = \left[
\begin{array}{cc}
		e^{(\boldsymbol{\sigma}/{2})\cdot\boldsymbol{\varphi}} & \0\\
		\0 & e^{-(\boldsymbol{\sigma}/{2})\cdot\boldsymbol{\varphi} }
\end{array}
\right] \nonumber\\
& = \sqrt{\frac{E + m }{2 m}}
						\left[
						\begin{array}{cc}
						\I + \frac{\boldsymbol{\sigma}\cdot\mathbf{p}}{E +m} & \0 \\
						\0 & \I - \frac{\boldsymbol{\sigma}\cdot\mathbf{p}}{E +m} 
						\end{array}
						\right].  \label{eq:sb}
\end{align}

To provide a concrete example of a mass dimension one quantum field, we confine our attention to  
the $\lambda(k^\mu)$-defining $\phi_L(k^\mu)$ as eigenspinors of  
$\s\cdot\widehat \p$ -- the helicity operator, modulo a factor of $\frac12$
\begin{equation}
\s\cdot\widehat \p\, \phi_L^\pm(k^\mu) = \pm \phi_L^\pm(k^\mu).\label{eq:h}
\end{equation}
 We adopt the `locality phases' so that
\begin{align}
\phi^+_L(k^\mu) & = \sqrt{m} \left[
									\begin{array}{c}
									\cos(\theta/2)\exp(- i \phi/2)\\
									\sin(\theta/2)\exp(+i \phi/2)
											\end{array}
									\right] 
									 =  \phi^+_L(\bf{0}) \Big\vert_{ \mathrm{Eq.\;(A.2)\;of\;AG} }\label{eq:zimpok52} \\
\phi^-_L(k^\mu) & = \sqrt{m} \left[
									\begin{array}{c}
									-\sin(\theta/2)\exp(- i \phi/2)\\
									\cos(\theta/2)\exp(+i \phi/2)
											\end{array}
									\right]
									=  - \phi^-_L(\bf{0}) \big \vert_{\mathrm{Eq.\;(A.3)\;of\;AG}} 
\end{align}
These differ from that in an earlier work~\cite{Ahluwalia:2004ab}, abbreviated above as AG.
There is a second choice of the  locality phases, and designations (that is, the indices the $\lambda_\alpha(k^\mu)$ are assigned), which is invoked when the $\lambda^{S,A}(p^\mu)$ are  used as the expansion coefficients of a quantum field. This choice we make explicit below in defining the $\lambda(k^\mu)$
\begin{align}
 \lambda^S_+(k^\mu) & = + \left[
					\begin{array}{c}
					i \Theta\left[\phi_L^+(k^\mu)\right]^\ast\\
								\phi_L^+(k^\mu) 
					\end{array}
					\right]  
					 =  +\,\lambda^S_{\{-,+\}}(\bf{0}) \Big 
					\vert_{\mathrm{Of\; (3.9)\; of\; AG}}   \label{eq:zimpok0}\\
\lambda^S_-(k^\mu) & = + \left[
				\begin{array}{c}
				i \Theta\left[\phi_L^-(k^\mu)\right]^\ast\\
				\phi_L^-(k^\mu)
				\end{array}
				\right] 
				= - \,\lambda^S_{\{+,-\}}(\bf{0}) 
				\Big \vert_{\mathrm{Of\; (3.9)\; of\; AG}}
\end{align}
and
\begin{align}
 \lambda^A_+(k^\mu) & = + \left[
					\begin{array}{c}
					- i \Theta\left[\phi_L^-(k^\mu)\right]^\ast\\
								\phi_L^-(k^\mu)
					\end{array}
					\right] 
					 = - \,\underbrace{\lambda^A_{\{+,-\}}(\bf{0})}_{\mathrm {and\; not\;}
				\lambda^A_{\{-,+\}}(\bf 0)}
												 \Big \vert_
												 {\mathrm{Of\; 		(3.10)\; of\; AG}} \label{eq:desg1}\\
	\lambda^A_-(k^\mu) &= - \left[
				\begin{array}{c}
				- i \Theta\left[\phi_L^+(k^\mu)\right]^\ast\\
					\phi_L^+(k^\mu)
				\end{array}
				\right]
				= - \,\underbrace{\lambda^A_{\{-,+\}}(\bf{0})}_{{\mathrm{ and\; not}\;}
										\lambda^A_{\{+,-\}}(\bf 0)}
												 \Big \vert_{\mathrm{Of\; (3.10)\; of\; AG}}\label{eq:zimpok91}
\end{align}

If one wishes one can keep the here-chosen phases free and fix them later by demanding locality for the resulting quantum field.

When the $\lambda(k^\mu)$ defined in (\ref{eq:zimpok0}) to (\ref{eq:zimpok91}) are acted upon by the boost (\ref{eq:sb}) we obtain a complete set of $\lambda(p^\mu)$. In carrying out this exercise a significant simplification occurs if one exploits the identity 
\begin{equation}
		\s\cdot\widehat{\p} \,
		\Big[ \Theta [\phi^\pm_L(k^\mu)]^\ast \Big] = \mp \Big[ \Theta [\phi^\pm_L(k^\mu)]^\ast \Big] .		\label{eq:zimpok3}
\end{equation}
It is deciphered by  complex conjugating (\ref{eq:h}), then replacing $\s^\ast$ in accord with~(\ref{eq:wt}), using  $\Theta^{-1} = -\Theta$, and finally multiplying from the left by $\Theta$. It reveals that helicity of $ \Theta [\phi_L(k^\mu)]^\ast$ is opposite to that of 
$\phi_L(k^\mu)$ -- precisely as indicated in (\ref{eq:zimpok3}).
The interplay of the result (\ref{eq:zimpok3}) with the boost
(\ref{eq:sb}) and the chosen form of  $\lambda(k^\mu)$ in (\ref{eq:zimpok0}) to 
 (\ref{eq:zimpok91}) gives the following form for $\lambda(p^\mu)$
\begin{align}
		\lambda^S_+(p^\mu) &= \sqrt{\frac{E+m}{2 m} }\left[ 1-\frac{p}{E+m}\right]\lambda^S_+(k^\mu),\nonumber \\		
		\lambda^S_-(p^\mu) &= \sqrt{\frac{E+m}{2 m} }\left[ 1+\frac{p}{E+m}\right]\lambda^S_-(k^\mu) 		\label{eq:lsm}
\end{align}
and
\begin{align}
		\lambda^A_+(p^\mu) &= \sqrt{\frac{E+m}{2 m} }\left[ 1+\frac{p}{E+m}\right]\lambda^A_+(k^\mu),\nonumber\\
		\lambda^A_-(p^\mu) &= \sqrt{\frac{E+m}{2 m} }\left[ 1-\frac{p}{E+m}\right]\lambda^A_-(k^\mu). 
		\label{eq:lam}
\end{align}
In sharp contrast to the eigenspinors of the parity operator~\textendash~that is, the Dirac spinors~\textendash~the here-considered eigenspinors of the charge conjugation operator, $\lambda^{S,A}_\pm(p^\mu)$, are simply the rest spinors $\lambda(k^\mu)$ scaled by the indicated energy-dependent factors. The boost does not mix various components of these `rest-frame' spinors. 

An inspection of (\ref{eq:lsm}) and (\ref{eq:lam}) suggests that for massless $\lambda(p^\mu)$ the number of degrees of freedom reduces to two, that is those associated with $\lambda^S_-(p^\mu)$ and $\lambda^A_+(p^\mu)$; $\lambda^S_+(p^\mu)$ and $\lambda^A_-(p^\mu)$ vanish identically. 

We will see below that
the parity operator takes $\lambda_-^{S}(p^\mu) \to \lambda_+^{S}(p^\mu)$ and $\lambda_+^{A}(p^\mu) \to \lambda_-^{A}(p^\mu)$. Combining these two observations we conclude that in the massless limit $\lambda(p^\mu)$ have no reflection.

Strictly speaking for massless particles there is no rest frame, or `rest-frame' spinors. The theory must be constructed \textit{ab initio} except that the massless limit 
of certain massive representation spaces yields the massless theory. 
The $(1/2,0)$ and $(0,1/2)$ representation spaces belongs to that class.
We refer the reader to Weinberg's 1964 work on the subject~\cite{PhysRev.134.B882}.
That such a limit may be taken for the $(1/2,0)\oplus(0,1/2)$ representation space is apparent from Weinberg's analysis.

The $\lambda^{S,A}_\pm(p^\mu)$ are the expansion coefficients of the new quantum field to be introduced below.

\subsection{The $\bm{\lambda(p^\mu)}$ do not satisfy Dirac equation: a hint towards mass dimension one fermions\label{sec:elko-do-not-satisfy-Dirac-equation}}

A hint for the unexpected theoretical discovery of spin-one-half fermions with mass dimension one  resides in the observation that the momentum-space Dirac operator $(\gamma_\mu p^\mu \pm m )$ does not annihilate the introduced eigenspinors of the charge conjugation operator. To establish this result, keeping (\ref{eq:lsm}) and (\ref{eq:diracgamma-lower}) in mind, we  operate $\gamma_\mu p^\mu$ on $\lambda^S_+(p^\mu)$:
\begin{align}
\gamma_\mu p^\mu   \lambda^S_+(p^\mu)  = \sqrt{\frac{E+m}{2 m}} & \left[ 1 - \frac{p}{E+m}\right]
\nonumber \\
  & \times \left[
  E  \gamma_0  + p \left(\begin{array}{cc}
 0 & \s\cdot\widehat{\p} \\
 -\s\cdot\widehat{\p} & 0
 \end{array}\right) 
  \right]  \lambda^S_+(k^\mu).\label{eq:zimpok4}
 \end{align}

To proceed further we note that 
\begin{align}
\left(\begin{array}{cc}
 0 & \s\cdot\widehat{\p} \\
 -\s\cdot\widehat{\p} & 0
 \end{array}\right) &\lambda^S_+(k^\mu) \nonumber\\
 &= \left(\begin{array}{cc}
 0 & \s\cdot\widehat{\p} \\
 -\s\cdot\widehat{\p} & 0
 \end{array}\right)   
 \left(
					\begin{array}{c}
					i \Theta\left[\phi_L^+(k^\mu)\right]^\ast\\
								\phi_L^+(k^\mu) 
					\end{array}
					\right)  .
\end{align}
But $\s\cdot\widehat{\p} \,\phi_L^+(k^\mu) 
 = \phi_L^+(k^\mu) $, while according to (\ref{eq:zimpok3})
 \begin{equation}\s\cdot\widehat{\p} \,\Big[\Theta\left[\phi_L^+(k^\mu)\right]^\ast\Big]
 = - \Theta\left[\phi_L^+(k^\mu)\right]^\ast.
 \end{equation}
Therefore, we have the result
\begin{align}
\left(\begin{array}{cc}
 0 & \s\cdot\hat{\p} \\
 -\s\cdot\hat{\p} & 0
 \end{array}\right)& \lambda^S_+(k^\mu) \nonumber\\
& = 
\left(\begin{array}{c}
\phi_L^+(k^\mu)\\
i \Theta\left[\phi_L^+(k^\mu)\right]^\ast
\end{array}
\right)  \nonumber\\
& = \left(\begin{array}{cc}
\0 & \openone \\
\openone & \0\end{array}\right) 
\left(\begin{array}{c}
i \Theta\left[\phi_L^+(k^\mu)\right]^\ast\\
\phi_L^+(k^\mu)
\end{array}
\right) \nonumber\\
& = \gamma_0 \lambda^S_+(k^\mu).
\end{align}

\noindent
As a consequence (\ref{eq:zimpok4}) simplifies   to 
\begin{equation}
  \gamma_\mu p^\mu  \lambda^S_+(p^\mu)   =    \sqrt{\frac{E+m}{2 m}} \left(1 - \frac{p}{E+m}\right)
\left(
  E    + p 
  \right)  \gamma_0\lambda^S_+(k^\mu).\label{eq:zimpok5}
 \end{equation}
The  standard dispersion relation allows for the replacement 
\begin{equation}
\left(1 - \frac{p}{E+m}\right)(E+p)
\rightarrow  m \left(1 + \frac{p}{E+m}\right)
\end{equation}
While on the other hand we have the identity
	\begin{equation}
		\gamma_0 \lambda^S_+(k^\mu) = i \lambda^S_-(k^\mu) .
	\end{equation} 
Combined, these two observations reduce (\ref{eq:zimpok5}) to
	\begin{equation}
  	\gamma_\mu p^\mu  \lambda^S_+(p^\mu)  =   i m \sqrt{\frac{E+m}{2 m}} 	\left(1 + \frac{p}{E+m}\right) \lambda^S_-(k^\mu).\label{eq:zimpok6}
 	\end{equation}
Using (\ref{eq:lsm}) in the right-hand side of (\ref{eq:zimpok6}) gives
 \begin{equation}
  \gamma_\mu p^\mu  \lambda^S_+(p^\mu) = i m  \lambda^S_-(p^\mu).    \label{eq:er-a1}
 \end{equation}
An exactly similar exercise complements (\ref{eq:er-a1}) with
\begin{align}
 \gamma_\mu p^\mu \lambda^S_-(p^\mu) & = - i m \lambda^S_+(p^\mu) \label{eq:er-a2}\\
 \gamma_\mu p^\mu \lambda^A_-(p^\mu) & = i m \lambda^A_+(p^\mu)  \label{eq:er-b1}\\
 \gamma_\mu p^\mu \lambda^A_+(p^\mu) & = - i m \lambda^A_-(p^\mu) . \label{eq:er-b2}
\end{align}

Thus  the result:  $(\gamma_\mu p^\mu \pm m)$ does not annihilate the eigenspinors of charge conjugation operator $\mathcal{C}$.  
Dvoeglazov~\cite{Dvoeglazov:1995eg,Dvoeglazov:1995kn} has shown that our earlier 1994 
preprint~\cite{Ahluwalia:1994uy}  already contained this result implicitly. To the best our knowledge the derivation provided here is the most direct and unambiguous calculation.

Equations (\ref{eq:er-a1}) to  (\ref{eq:er-b2}), coupled with the discussion surrounding  (\ref{eq:counterpart}), contain the rudimentary seeds for the kinematical and dynamical content of the quantum field built upon $\lambda(p^\mu)$ as its expansion coefficients: first, $\lambda(p^\mu)$ are annihilated by the spinorial Klein-Gordon operator (and not by the Dirac operator), and second, 
the resulting kinematic structure cannot support the usual gauge symmetries of the standard model of the high energy physics.
In regard to the  the former claim, we multiply (\ref{eq:er-a1}) from the left by $\gamma_\nu p^\nu $
\begin{align}
\gamma_\nu p^\nu 
\gamma_\mu p^\mu  \lambda^S_+(p^\mu)  &\stackrel{(\ref{eq:er-a1})}{=}  i m  \gamma_\nu p^\nu \lambda^S_-(p^\mu) \nonumber\\
&\stackrel{(\ref{eq:er-a2})}{=} im \left( - i m \lambda^S_+(p^\mu)\right) = m^2 \lambda^S_+(p^\mu)
\label{eq:zimpok1952}
\end{align}
and noting that  the left hand side of the above equation can be rewritten exploiting
$\{\gamma_\mu,\gamma_\nu\}= 2 \eta_{\mu\nu} \I_4$ (where $\eta_{\mu\nu}$ is the space-time metric with signature $(+1,-1,-1,-1)$)
\begin{align}
 \gamma_\nu p^\nu 
\gamma_\mu p^\mu & = \frac{1}{2}\left( \gamma_\nu p^\nu 
\gamma_\mu p^\mu  + \gamma_\mu p^\mu 
\gamma_\nu p^\nu \right) p^\mu p^\nu \nonumber\\
&= \frac{1}{2}\{\gamma_\mu,\;\gamma_\nu\}   p^\mu p^\nu 
 = \eta_{\mu\nu}p^\mu  p^\nu \I_4 .
\end{align}
Substituting this result in~(\ref{eq:zimpok1952}), and rearranging gives
\begin{equation}
(\eta_{\mu\nu}p^\mu  p^\nu \I_4 - m^2\I_4)\lambda^{S}_+(p^\mu) = 0.
\end{equation}
Repeating the same exercise with  (\ref{eq:er-a2}) to  (\ref{eq:er-b2}) as the starting point, yields
\begin{equation}
(\eta_{\mu\nu}p^\mu  p^\nu \I_4 - m^2\I_4)\lambda^{S,A}_\pm(p^\mu) = 0. \label{eq:skg}
\end{equation}

\subsubsection{$\bm{\mathcal{CPT}}$ and Elko\label{sec:parenthetic-observation}}

In conjunction with (\ref{eq:P}), equations (\ref{eq:er-a1})-(\ref{eq:er-b2}) also serve to yield the action of $\mathcal{P}$ on $\lambda_\pm^{\mathrm{S,A}}(p^\mu)$
 \begin{align}
  \mathcal{P} \lambda^S_+(p^\mu) &= i   \lambda^S_-(p^\mu)   \label{eq:er-a1P}\\
    \mathcal{P} \lambda^S_-(p^\mu) & = - i  \lambda^S_+(p^\mu) \label{eq:er-a2P}\\
   \mathcal{P}  \lambda^A_-(p^\mu) & = i  \lambda^A_+(p^\mu)  \label{eq:er-b1P}\\
  \mathcal{P}  \lambda^A_+(p^\mu) & = - i  \lambda^A_-(p^\mu) . \label{eq:er-b2P}
\end{align}
These can be compacted into the following
\begin{equation}
  \mathcal{P} \lambda^S_\pm (p^\mu) = \pm i   \lambda^S_\mp(p^\mu),\quad
 \mathcal{P}\lambda^A_\pm (p^\mu)  = \mp i  \lambda^A_\mp(p^\mu) \label{eq:parity-b2}
\end{equation}
and lead to $\mathcal{P}^2=\I_4$.
Acting $\mathcal{C}$ from the left on the first of the above equations gives
\begin{equation}
\mathcal{C} \mathcal{P} \lambda^S_+(p^\mu) = - i \mathcal{C}   \lambda^S_-(p^\mu) =  -i \lambda^S_-(p^\mu) . \label{eq:zimpok100}
\end{equation}
On the other hand
\begin{equation}
\mathcal{P} \mathcal{C}  \lambda^S_+(p^\mu) =
\mathcal{P}  \lambda^S_+(p^\mu) = i   \lambda^S_-(p^\mu) .
\end{equation}
Adding the above two results leads to anti-commutativity  for the $\mathcal{C}$ and $\mathcal{P}$ for $ \lambda^S_+(p^\mu) $.
Repeating the same exercise for $\lambda^S_-(p^\mu)$ and $\lambda^A_\pm(p^\mu) $
 establishes
that $\mathcal{C}$ and $\mathcal{P}$ anticommute for all $\lambda(p^\mu)$:
\begin{equation}
\{\mathcal{C},\;\mathcal{P}\} = 0\label{eq:cp-z}
\end{equation}

The time-reversal operator $\mathcal{T} = i \gamma \mathcal{C}$ acts on Elko as follows
\begin{equation}
\mathcal{T} \lambda^S_\alpha(p^\mu) = - i \lambda^A_\alpha(p^\mu),\quad
\mathcal{T} \lambda^A_\alpha(p^\mu) = +i \lambda^S_\alpha(p^\mu),
\end{equation}
For completeness, we note the counterparts of (\ref{eq:cp-z})
\begin{equation}
\left[\mathcal{T},\mathcal{P}\right]=0,\quad \left[\mathcal{C},\mathcal{T}\right]=0
\end{equation}
As a  consequence 
\begin{equation}
\left(\mathcal{C}\mathcal{P}\mathcal{T}\right)^2 = + \I_4.\label{eq:cpt-sq}
\end{equation}


\vspace{9pt}

 For the $\mathfrak{m}(x)$ field introduced in section~\ref{sec:Majorana-isation} an
internal consistency for the transformation of the creation and annihilation operators under parity demands to incorporate the phase 
$\pm\exp(i\pi/2)$ in the definition of $\mathcal{P}$. It leads to commutativity of $\mathcal{P}$ and
$\mathcal{C}$: 
\begin{equation}
[\mathcal{C},\;\mathcal{P}] = 0
\end{equation} 
with 
$\mathcal{P}^2=-\I_4$.\footnote{I acknowledge correspondence with Otto Nachtmann and Marek Zralek on the subject when the present work was still to be formulated.}
We still have $\mathcal{C}^2 =\I_4$ and $\mathcal{T}^2=-\I_4$. Again $\mathcal{C}\mathcal{P}\mathcal{T}$ ceases to be an identity, but instead we have 
\begin{equation}
\left(\mathcal{C}\mathcal{P}\mathcal{T}\right)^2 = - \I_4\label{eq:cpt-b}
\end{equation}

\noindent
\textbf{A parenthetic remark.}  All commutators and anticommutators are understood to  hold true while acting on $\lambda_\alpha(p^\mu)$. Some of the results presented in this section differ from those presented in our earlier publications, and those by Wunderle and Dick~\cite{Wunderle:2009zz}. It is because $\mathcal{P}$, as presented  in equation (\ref{eq:P}), removes the ambiguities regarding the action of the parity on the helicity indices.

\section{The mathematical theory of spinorial duals\label{sec:the-mathematical-theory-of-duals} }

The result that the spin one-half eigenspinors of the charge conjugation operator satisfy only the Klein Gordon equation would suggest that at the `classical level' the Lagrangian density for the 
$\lambda(x)$ would simply be
\begin{equation}
\mathfrak{L}(x) = \partial^\mu{\overline{\lambda}(x)}\,\partial_\mu {{\lambda(x)}} - m^2 {\overline{\lambda}}(x) \lambda(x)\label{eq:wrong}
\end{equation}
where $\lambda(x)$ is a classical field with $\lambda^{S,A}(p^\mu)$ as its Fourier coefficients.
This  apparently natural choice is too naive. Its validity is challenged by  an explicit calculation which shows that under the Dirac dual
\begin{equation}
\overline{\lambda}^{S,A}_\alpha(p^\mu)\stackrel{\rm def}{=}\left[\lambda^{S,A}_\alpha(p^\mu)\right]^\dagger \gamma_0 .
\end{equation}
the norm of each of the four $\mathcal{C}$ eigenspinors $\lambda^{S,A}_\alpha(p^\mu)$ identically vanishes
\begin{equation}
\overline{\lambda}^S_\alpha(p^\mu) \lambda^S_\alpha(p^\mu) = 0,\quad
\overline{\lambda}^A_\alpha(p^\mu) \lambda^A_\alpha(p^\mu) = 0 .
\end{equation}

\subsection{A temptation, and a departure\label{sec:temptation}}

One may thus be tempted to suggest that we introduce, instead, a Majorana mass term and treat the components of $\lambda(p^\mu)$ as anticommuting numbers (that is, as Grassmann numbers).\footnote{A pedagogic introduction to which may be found in the Lancaster and Blundell's 2014 QFT book~\cite[Chapter 48]{LancasterBlundell:2014tsj}.
A complimentary and more detailed treatment can be found in Matthew Schwartz 
monograph~\cite{Schwartz:2014md}.
} This, in effect, would immediately promote a c-number classical field to the quantum field of Majorana and demand that the kinetic term be restored to that of Dirac. In the process we will be forced to abandon Elko as possible expansion coefficients of a quantum field, and return to a field with Dirac spinors as its expansion coefficients. The resulting field would then have the form of the Majorana field~\cite{Majorana:1937vz}
	\begin{equation}
	 \left(
	\begin{array}{c}
	i \Theta \Psi^\ast_L(x)\\
	\Psi_L(x)
	\end{array}
	\right) 
	\end{equation} 
where $\Psi_L(x)$ is the left-transforming projection of the $\Psi(x)$, the Dirac field with  the particle and antiparticle creation operators identified with each other.

If we do not fall into this temptation: an unexpected theoretical result follows that naturally leads us to a new class of fermions of spin one half. 

\subsection{The questions on the path of departure\label{sec:The-questions-on-the-path-of-departure}}

On our path of departure,  some of the questions are 
  \begin{quote}
1.  If the Dirac dual
\begin{equation}\overline{\psi}(p^\mu)=\left[\psi(p^\mu)\right]^\dagger\gamma_0 \label{eq:Dirac-dual}
\end{equation}
was not given how shall we go about deciphering it?\\\noindent
2.  Is (\ref{eq:Dirac-dual}) a unique dual, or is there a freedom in its definition?, and what physics does it encode?
\end{quote}
These questions, if ever, to the best of our knowledge, are rarely asked in the physics literature. An elegant exception for the definition, called a convenience by Weinberg, is to note, following Weinberg, that the counterpart of $\Lambda$ in (\ref{eq:Minkowski-boost-rotation}) for the $(1/2,0)\oplus(0,1/2)$ representation space
\begin{equation}
D(\Lambda) \stackrel{\mathrm{def.}}{=}\left\{
\begin{array}{cl}
\exp\left(i\kb\cdot\vp\right) & \mbox{for Lorentz boosts} \\
\exp\left(i\bz\cdot\vp\right) & \mbox{for rotations}
\end{array}\right.
\end{equation}
is not unitary, but pseudounitary\footnote{\label{fn:8}Strictly speaking, Weinberg's discussion is for a   spin one half quantum field
 but it readily adapts to the c-number spinors, $\psi(p^\mu)$, in the $(1/2,0)\oplus(0,1/2)$ representation space.}
\begin{equation}
\gamma_0 D(\Lambda)^\dagger \gamma_0 = D(\Lambda)^{-1}\label{eq:pseudounitarity}
\end{equation}
Weinberg uses this observation to motivate the definition (\ref{eq:Dirac-dual}) -- 
however, see footnote~\ref{fn:8}.

\subsection{A need for a new dual for Elko\label{sec:A-need-for-a-new-dual-for-Elko}}

Now for Elko  ($m\neq 0$), the Dirac dual 
\begin{equation}
\overline\lambda(p^\mu)\stackrel{\mathrm{def}}{=} \lambda(p^\mu)^\dagger \gamma_0 .
\end{equation}
yields a   null norm~\cite{Ahluwalia:1994uy}:
\begin{align}
	&\overline\lambda^S_\pm(p^\mu)  \lambda^S_\pm(p^\mu)  = 0,  \quad \overline\lambda^S_\pm(p^\mu)	\lambda^A_\pm(p^\mu) = 0,\quad 
	 \overline\lambda^S_\pm(p^\mu) \lambda^A_\mp(p^\mu) =0 \label{eq:norm-a}\\
	 &\overline\lambda^A_\pm(p^\mu) \lambda^A_\pm(p^\mu)  = 0,\quad \bar\lambda^A_\pm(p^\mu) \lambda^S_\pm(p^\mu) = 0,\quad
	 \overline\lambda^A_\pm(p^\mu) \lambda^S_\mp(p^\mu) =0. \label{eq:norm-b} 
\end{align}
But because
\begin{equation}
\overline\lambda^S_\pm(p^\mu) \lambda^S_\mp(p^\mu) = \mp \,2 i m,\quad
	 \overline\lambda^A_\pm(p^\mu) \lambda^A_\mp(p^\mu)  = \pm \,2 i m \label{eq:norm-c}
	 \end{equation}
one can define  a new dual and make it convenient for oneself to formulate and calculate the physics of
quantum fields with Elko as their expansion coefficients. Once a lack of uniqueness of the Dirac dual is discovered, the new way to accommodate the pseudounitarity 
captured by (\ref{eq:pseudounitarity}) is introduced.  And it opens up concrete new possibilities to go beyond the 
Dirac and Majorana fields for spin one half fermions without violating Lorentz covariance and without introducing non-locality.

\subsection{The dual of spinors: \underline{constraints from the scalar invariants} \label{sec:constraint-from-scalar-invariants}}

Consider a general set of $4$-component massive spinors $\varrho(p^\mu)$ -- we would like these to be orthonormal under the dual we are seeking. These do not have to be  eigenspinors of $\mathcal{P}$, or eigenspinors of $\mathcal{C}$. We examine a general form of the dual defined  as
\begin{equation}
\dual{\varrho}_\alpha(p^\mu) \stackrel{\rm def}{=}{\big[}\Xi(p^\mu) \, \varrho_\alpha(p^\mu){\big]}^\dagger \eta .
\label{eq:gendual}
\end{equation}
The $\Xi(p^\mu)$ and $\eta$ are $4\times 4$ matrices, with the elements of $\eta$ in $\mathbb{C}$.  

The task of $\Xi(p^\mu)$  is to take any one of the $\varrho_\alpha(p^\mu)$ and transform it, up to a phase,
 into one of the spinors $\varrho_{\alpha^\prime}(p^\mu)$ from the same set. It is not necessary that the indices $\alpha^\prime$ and $\alpha$ be the same. 
 We require $\Xi(p^\mu)$ to define an invertible map, with $\Xi^2 = \I_4$ (possibly, up to a phase). 
 \vspace{11pt}
 
 \noindent
 \textbf{Two examples for $\bm{\Xi(p^\mu)}$}
 
 For example, for 
\begin{equation}\Xi(p^\mu) =\I_4\label{sec:Dirac-Xi}
\end{equation}
we have a simple map
 \begin{equation}
 \varrho_\alpha(p^\mu)\to\varrho_\alpha(p^\mu),\quad\forall\alpha
 \end{equation} 
 While, keeping the results  (\ref{eq:norm-a}) to (\ref{eq:norm-c}) in mind, the choice~\cite{Speranca:2013hqa}
 \begin{align}
	\Xi(p^\mu)   \stackrel{\rm def}{=}  \frac{1}{2 m} \sum_{\alpha=\pm}	
			  \Big[\lambda^S_\alpha(p^\mu)\bar\lambda^S_\alpha(p^\mu) 	 - \lambda^A_\alpha(p^\mu)\bar\lambda^A_\alpha(p^\mu) 	\Big],\label{eq:Xi}
	\end{align}
induces the following map if $\rho_\alpha(p^\mu)$ are taken as $\lambda_\alpha(p^\mu)$
\begin{align}
\lambda^S_+(p^\mu)& \to i \lambda^S_-(p^\mu)\label{eq:map1}\\
\lambda^S_-(p^\mu)& \to-  i \lambda^S_+(p^\mu)\label{eq:map2}\\
\lambda^A_+(p^\mu)& \to - i \lambda^A_-(p^\mu)\label{eq:map3}\\
\lambda^A_-(p^\mu)& \to+  i \lambda^S_+(p^\mu).\label{eq:map4}
\end{align}

We now wish to determine the metric $\eta$, and $\Xi(p^\mu)$ -- formally (we will see that the examples we have chosen for $\Xi(p^\mu)$ are indeed allowed).
For the boosts, the requirement of a Lorentz invariant norm translates to
\begin{equation}
\Big[\Xi(k^\mu)\varrho(k^\mu)\Big]^\dagger  \eta \,\varrho(k^\mu)  
= \Big[\Xi(p^\mu)\varrho(p^\mu)\Big]^\dagger \eta \,\varrho(p^\mu) 
\label{eq:boostdemand}
\end{equation}
 with a similar expression for the rotations. 
Expressing $\varrho(p^\mu)$ as  $\exp(i \kb\cdot\vp) \varrho(k^\mu)$, and using $\kb^\dagger = -\kb$ (for an explicit form of  $\bm{\kappa}$ see Eq.~(\ref{eq:pi})) ,
the right-hand side of the above expression can be re-written as 
\begin{equation}
\Big[\Xi(p^\mu)\varrho(p^\mu)\Big]^\dagger \eta \,\varrho(p^\mu) 
= \varrho^\dagger(k^\mu)\, e^{i\boldsymbol{\kappa\cdot\varphi}} \, \Xi^\dagger(p^\mu)\, \eta\, e^{i\boldsymbol{\kappa\cdot\varphi}}\, \varrho(k^\mu).\label{eq:breq}
\end{equation}
Using 
\begin{equation}
\Xi(p^\mu) = e^{i\boldsymbol{\kappa\cdot\varphi}}\, \Xi(k^\mu)\, e^{-i\boldsymbol{\kappa\cdot\varphi}} \label{eq:XiTransformation-b}
\end{equation}
reduces (\ref{eq:breq}) to
 	\begin{equation}
	\Big[\Xi(p^\mu)\varrho(p^\mu)\Big]^\dagger \eta \,\varrho(p^\mu) 
 =
 	\Big[\Xi(k^\mu) \varrho(k^\mu)\Big]^\dagger e^{i\boldsymbol{\kappa\cdot\varphi}} \eta \,e^{i\boldsymbol{\kappa\cdot\varphi}}\varrho(k^\mu).
 	\end{equation}
Comparing the above expression with  Eq.~(\ref{eq:boostdemand}) gives the constraint 
\begin{equation}
e^{i\boldsymbol{\kappa\cdot\varphi}} \eta \,e^{i\boldsymbol{\kappa\cdot\varphi}} = \eta .
\end{equation}
That is, the metric $\eta$ must anticommute with 
each of boost generators $\kb$ 	
\begin{equation}
	\{\kappa_i,\eta\}= 0,\qquad i = x,y,z.  	 \label{eq:eta-boost}
	\end{equation}
Different choices for $\Xi(p^\mu)$ result is different spinorial duals.

\subsubsection{The Dirac dual\label{sec:Dirac-dual}}

The simplest choice for $\Xi(p^\mu)$ is the  identity operator as chosen
in (\ref{sec:Dirac-Xi}). We will now show that the well-known Dirac dual corresponds to this choice. With  
$\Xi(p^\mu) = \openone_4$, the counterpart 
of (\ref{eq:boostdemand})
for rotations reads
\begin{equation}
\varrho^\dagger(p_\mu) \, \eta \,\varrho(p_\mu)  
= \varrho^\dagger(p^\prime_\mu)\, \eta \,\varrho(p^\prime_\mu) 
\label{eq:rotationdemand}
\end{equation}
where $\varrho(p^\prime_\mu) = e^{i\bm{\zeta}\cdot\bm{\theta}}\varrho(p_\mu)$, and using $\bm{\zeta}^\dagger
=\bm{\zeta}$ (for an explicit form of  $\bm{\zeta}$ see Eq.~(\ref{eq:pi})) translates the right-hand side of the above equation 
to
\begin{equation}
\varrho^\dagger(p_\mu) \, e^{-i \bm{\zeta}\cdot{\bm{\theta}}}\, \eta 
e^{i \bm{\zeta}\cdot{\bm{\theta}}}\,\varrho(p_\mu).
\end{equation}
Comparing the above expression with  (\ref{eq:rotationdemand}) gives the constraint 
\begin{equation}
e^{-i \bm{\zeta}\cdot{\bm{\theta}}}\, \eta 
e^{i \bm{\zeta}\cdot{\bm{\theta}}} = \eta.
\end{equation}
That is, the metric $\eta$ must commute with 
each of rotation generators $\bm{\zeta}$ 
\begin{equation}
\left[{\zeta}_i,\eta\right] =0,\qquad i = x,y,z.  	 \label{eq:eta-rotation}
\end{equation}
The constraint that $\eta$ must anticommute with $\kb_i$ and commute with 
$\bm{\zeta}_i$ -- with the additional reality condition on the norm -- restricts the metric $\eta$ to have the form
\begin{equation}
\eta = \left(\begin{array}{cccc}
0 & 0 & a & 0\\
0 & 0 & 0 & a \\
b & 0 & 0 & 0\\
0 & b & 0 &0
\end{array}\right),\qquad a,b \in \Re .\label{eta-zimpok9}
\end{equation}
Following a method similar to the one used above, the demand for the norm to be invariant under the parity transformation $\mathcal{P}$ is obtained to be
\begin{equation}
\left[m^{-2} \left(\gamma^\mu p_\mu\right)^\dagger \,\gamma^0\eta\gamma^0\,\gamma^\mu p_\mu\right] - \eta =0. \label{eq:parityconstraint}
\end{equation}
It requires $b=a$, which finally reduces $\eta$ to 
\begin{equation}
\eta = a \gamma^0.
\end{equation} 
Convention can now be invoked to set $a=1$, giving the canonical Dirac dual: it is defined by the choice of $\Xi(p^\mu)=\openone_4$, and the constraints on $\eta$ given by (\ref{eq:eta-boost}), (\ref{eq:eta-rotation}
) and (\ref{eq:parityconstraint}).  

On the path of our departure we learn that the 
Dirac dual `naively' defined in section \ref{sec:The-questions-on-the-path-of-departure} has additional underlying structure. In particular, it gives us a choice to violate parity, or to preserve it, depending on whether we choose $a/b$ in $\eta$ of (\ref{eta-zimpok9}) to be unity, or different from unity. With  the former choice we reproduce the standard result (\ref{eq:Dirac-dual}), while the latter choice gives us a first-principle control on the extent to which parity may be violated in nature, or in a given physical process.

 \subsubsection{The dual for $\bm{\lambda(p^\mu)}$\label{sec:elko-dual}}

The  dual for $\lambda(p^\mu)$ first  introduced in Ref.~\cite{Ahluwalia:2003jt}, and refined in Ref.~\cite{Ahluwalia:2008xi,Ahluwalia:2009rh}. It can now be more systematically understood by observing that those results correspond to the     
$\Xi(p^\mu)$ of the second example considered above (and given in equation (\ref{eq:Xi})).
Expression (\ref{eq:Xi}) can be evaluated to yield a compact form
	\begin{equation}
	\Xi(p^\mu) =
	m^{-1}  \mathcal{G}(p^\mu) \gamma_\mu p^\mu.
	\label{eq:XizZmpok}
	\end{equation}
where $\mathcal{G}(p^\mu)$ is defined as
\begin{equation}
\mathcal{G}(p^\mu)  \stackrel{\mathrm{def}}{=}
\left(\begin{array}{cccc}
			0 & 0 & 0 & -i e^{-i\phi} \\
			0 & 0 & i  e^{i \phi} & 0 \\
			0 & -i  e^{-i \phi} & 0 & 0\\
			i  e^{i\phi} & 0 & 0 & 0
			\end{array}\right). \label{eq:G}
\end{equation}
Out of the four variables $m$, $p$, $\theta$ and $\phi$ that enter 
\[p^\mu = (E, p \sin\theta\cos\phi,p\sin\theta\sin\phi,p\cos\theta)\]
$\mathcal{G}(p^\mu) $ depends only on $\phi$. 
The analysis of the boost constraint remains unaltered, with the result that we still have
	\begin{equation}
	\{\kb_i,\eta\}= 0,\quad i = x,y,z . 	 \label{eq:eta-boost-b}
	\end{equation}
The analysis for the rotation constraint changes. It begins as
\begin{equation}
\Big[m^{-1}  \mathcal{G}(p^\mu)  \gamma^\mu p_\mu  \, \lambda(p_\mu)\Big]^\dagger \,\eta\, \lambda(p_\mu)
 =
\Big[m^{-1}  \mathcal{G}(p^{\prime\mu}) \gamma^\mu p^\prime_\mu \,  \lambda(p^\prime_\mu)\Big]^\dagger \,\eta\, \lambda(p^\prime_\mu)
\end{equation}
where $\lambda(p_\mu)$ represents any of the four $\lambda^{S,A}_\pm(p^\mu)$, and the primed quantities refer to their rotation-induced counterparts. The above expression simplifies on using the following identities
\begin{equation}
\mathcal{G}(p^\mu)\, \lambda(p_\mu) = \pm \, \lambda(p_\mu),\qquad [\mathcal{G}(p^\mu),\,\gamma^\mu p_\mu ]=0,\label{eq:important}
\end{equation}
where the upper sign in the first equation above holds for $\lambda^S(p_\mu)$ and the lower sign is for $\lambda^A(p_\mu)$. The result is
\begin{equation}
\Big[ \gamma^\mu p_\mu  \, \lambda(p_\mu)\Big]^\dagger \,\eta\, \lambda(p_\mu)
 =
\Big[\gamma^\mu p^\prime_\mu \,  \lambda(p^\prime_\mu)\Big]^\dagger \,\eta\, \lambda(p^\prime_\mu).\label{eq:rotation-constraint-b}
\end{equation}
Expressing $\lambda(p^{\prime\mu})$ as  $e^{i \bm{\zeta}\cdot\bm{\theta}} \lambda(p^\mu)$, and using $\bm{\zeta}^\dagger
=\bm{\zeta}$,
the right-hand side of the above expression can be written as 
\begin{equation}
\Big[\gamma^\mu p^\prime_\mu \,  \lambda(p^\prime_\mu)\Big]^\dagger \,\eta\, \lambda(p^\prime_\mu) =
\lambda^\dagger(p_\mu)\,
e^{-i\bm{\zeta}\cdot\bm{\theta}} 
\left(\gamma^\mu p^\prime_\mu \right)^\dagger\,\eta\,
e^{i\bm{\zeta}\cdot\bm{\theta}} \,\lambda(p_\mu). \label{eq:extra}
\end{equation}
On taking note that
\begin{equation}
\gamma^\mu p^\prime_\mu = e^{i\bm{\zeta}\cdot\bm{\theta}}  \gamma^\mu p_\mu 
e^{-i\bm{\zeta}\cdot\bm{\theta}}
\end{equation}
equation (\ref{eq:extra}) becomes
\begin{equation}
\Big[\gamma^\mu p^\prime_\mu \,  \lambda(p^\prime_\mu)\Big]^\dagger \,\eta\, \lambda(p^\prime_\mu) =
\Big[\gamma^\mu p_\mu\,\lambda(p_\mu)\Big]^\dagger\,
e^{-i\bm{\zeta}\cdot\bm{\theta}} 
\eta\,
e^{i\bm{\zeta}\cdot\bm{\theta}} \,\lambda(p_\mu)
\end{equation}
Comparing the above expression with  (\ref{eq:rotation-constraint-b}) gives the constraint 
\begin{equation}
e^{-i\bm{\zeta}\cdot\bm{\theta}} 
\eta\,
e^{i\bm{\zeta}\cdot\bm{\theta}}  = \eta
\end{equation}
That is, the metric $\eta$ must commute with 
each of rotation generators $\bm{\zeta}$ 
\begin{equation}
  \left[\zeta_i,\eta\right] = 0 ,\quad i = x,y,z  	 \label{eq:eta-rotation-c}
\end{equation}
Without a surprise, this is the same result as before, despite a non-trivial $\Xi(p^\mu)$.


It is readily seen that  $\left[\Xi(p^\mu)\right]^2=\I$ and $\left[\Xi(p^\mu)\right]^{-1}$ indeed  exists and equals $\Xi(p^\mu)$ itself. 
Thus, the
  dual for $\lambda(p^\mu)$ is defined by the choice of $\Xi(p^\mu)$ given by 
 (\ref{eq:XizZmpok}). 
 
 So far the constraints on $\eta$ turn out to be same as for the Dirac dual. 
 To distinguish it form other possibilities the new dual at the intermediate state of our calculations is represented by 
 \begin{equation}
\dual{\lambda}_\alpha(p^\mu)   \stackrel{\rm def}{=} {\big[}\Xi(p^\mu)\, \lambda_\alpha(p^\mu){\big]}^\dagger \eta
\label{eq:dual-b}
\end{equation}
with $\eta$ obtained by setting $a=b=1$ in (\ref{eta-zimpok9}). This choice is purely for convenience at the moment, and if the new particles to be introduced here are indeed an element of the physical reality then the ratio $a/b$ must not be set to unity but determined by appropriate observations/experiments.
 
The definition (\ref{eq:dual-b}) allows us to rewrite results (\ref{eq:norm-a}), (\ref{eq:norm-b}), and (\ref{eq:norm-c}) into the following orthonormality relations
\begin{align}
& \dual\lambda^S_\alpha(p^\mu) \lambda^S_{\alpha^\prime}(p^\mu)
 = 2 m \delta_{\alpha\alpha^\prime}\label{eq:zimpokJ9a}\\
&  \dual\lambda^A_\alpha(p^\mu) \lambda^A_{\alpha^\prime}(p^\mu)
 = - 2 m \delta_{\alpha\alpha^\prime} \label{eq:zimpokJ9b}\\
 & \dual\lambda^S_\alpha(p^\mu) \lambda^A_{\alpha^\prime}(p^\mu) = 0 =
 \dual\lambda^A_\alpha(p^\mu) \lambda^S_{\alpha^\prime}(p^\mu).\label{eq:zimpokJ9c}
\end{align}

\subsubsection{Equivalent representations of the same dual\label{sec:equivalent-representations}}
The map (\ref{eq:map1}-\ref{eq:map4}) can be summarised into
\begin{align}
\Xi(p^\mu) \lambda^S_\pm(p^\mu) = \pm i  \lambda^S_\mp(p^\mu)\\
\Xi(p^\mu) \lambda^A_\pm(p^\mu) = \mp i  \lambda^A_\mp(p^\mu).
\end{align}
The above map in conjunction with the definition (\ref{eq:dual-b}) readily translates to
\begin{align}
\dual\lambda^S_+(p^\mu) &= - i \left[\lambda^S_-(p^\mu)\right]^\dagger \eta  \label{eq:lspd} \\
\dual\lambda^S_-(p^\mu) &= i \left[\lambda^S_+(p^\mu)\right]^\dagger\eta\label{eq:otherd-1}\\
 \dual\lambda^A_+(p^\mu) &= -i \left[\lambda^A_-(p^\mu)\right]^\dagger\eta\label{eq:otherd-2}\\
 \dual\lambda^A_-(p^\mu) &= i \left[\lambda^A_+(p^\mu)\right]^\dagger\eta
\label{eq:otherd-3}
\end{align}

A comparison of these results with those given in~\cite[Eq. 15]{Ahluwalia:2008xi} and \cite[Eq. 22]{Ahluwalia:2009rh} establishes the equivalence of the  dual introduced here and the one introduced in the previous works. The new way of defining and obtaining the spinorial duals provides a justification for the earlier definitions and renders many of the calculations much simpler.

For comparison with the work presented in Refs.~\cite{Ahluwalia:2004sz,Ahluwalia:2004ab} we introduce
\begin{equation}
{\epsilon_{\alpha}}^\beta \stackrel{\rm def}{=} \left(
\begin{array}{cc} 
0 & -1 
\\ 1 & 0
\end{array}\right)
\end{equation}
with rows and columns marked as $\{+,-\}$. That done, one immediately sees that the new dual and the Dirac dual are related by the expression
\begin{equation}
\dual\lambda_\alpha(p^\mu) = i {\epsilon_{\alpha}}^\beta \lambda_\beta^\dagger(p^\mu) \eta = 
 i {\epsilon_{\alpha}}^\beta
 \overline{\lambda}_\beta(p^\mu)
 \label{eq:elkodual3}
\end{equation}
where the self/anti-self conjugacy indices $S$ and $A$ have been suppressed.
In the above expression a sum on $\beta$ is implicit, and the position of the indices on $\epsilon$ is for convenience and it is not intended that these indices are raised and lowered using a metric.

Definitions (\ref{eq:dual-b}), $\{$(\ref{eq:lspd}), (\ref{eq:otherd-1}-\ref{eq:otherd-3})$\}$,
and (\ref{eq:elkodual3}) are three different, but equivalent, representations of dual for 
$\lambda(p^\mu)$.

\subsection{The dual of spinors: \underline{constraint from the invariance of the Elko spin sums}\label{sec:constraints-from-the-spin-sums}}

The Dirac spin sums place no additional constraints on the Dirac dual and we thus omit those calculations and confine to the $\mathcal{C}$ eigenspinors (that is, Elko).

The spin sums for the eigenspinors of the spin one-half charge conjugation operator $\mathcal{C}$ under the introduced dual
\begin{equation}
\sum_{\alpha} \lambda^S_\alpha(p^\mu) \dual \lambda^S_\alpha(p^\mu) \;\;{\rm and} \;\;
\sum_{\alpha}\lambda^A_\alpha(p^\mu) \dual \lambda^A_\alpha(p^\mu) 
\end{equation}
can now be readily evaluated
using Eqs.~(\ref{eq:lsm}-\ref{eq:lam}) for the $\lambda^S_\alpha(p^\mu)$ and
 $\lambda^A_\alpha(p^\mu)$,   and equations~(\ref{eq:lspd}-\ref{eq:otherd-3})$\}$ for their duals. The first of the two spin sums evaluates to
\begin{equation}
 	  i\; \underbrace{ \left[ \frac{E + m}{2 m}
 	 \left(1 - \frac{{p^2}}{(E+m)^2}\right)\right]}_{=1}    		\underbrace{\bigg( - \lambda^S_+(k^\mu) \left[\lambda^S_-(k^\mu)		\right]^\dagger
  			+ \lambda^S_-(k^\mu) \left[\lambda^S_+(k^\mu)\right]^\dagger
			\bigg)\eta}_{= -i m \left[ \I_4 + \mathcal{G}(p^\mu) \right]}			\nonumber
\end{equation}
The second of the spin sums can be evaluated in exactly the same manner. The combined result is
\begin{align}
\sum_{\alpha} \lambda^S_\alpha(p^\mu) \dual \lambda^S_\alpha(p^\mu) & = +
m \big[\I_4 + \mathcal{G}(p^\mu) \big] \label{eq:sss}\\
\sum_{\alpha} \lambda^A_\alpha(p^\mu) \dual \lambda^A_\alpha(p^\mu) & =
- m \big[\I_4 - \mathcal{G}(p^\mu)\big]  \label{eq:ssa}
\end{align}
with $\mathcal{G}(p^\mu)$ as in Eq.~(\ref{eq:G}).

These spin sums have the eigenvalues $\{0,0,2m,2m\}$, and $\{0,0,-2m,-2m\}$, respectively. Since eigenvalues  of projectors  must be either zero or one~\cite[Section 3.3]{Weinberg:2012qm}, we define
\begin{align}
  P_S &\stackrel{\mathrm{def}}{=}\frac{1}{2 m} \sum_{\alpha} \lambda^S_\alpha(p^\mu) \dual \lambda^S_\alpha(p^\mu)=  \frac{1}{2} \big[\I_4 + \mathcal{G}(p^\mu) \big]  \label{eq:s}
\\
P_{A}&\stackrel{\mathrm{def}}{=} -\frac{1}{2 m} \sum_{\alpha} \lambda^A_\alpha(p^\mu) \dual \lambda^A_\alpha(p^\mu) =  \frac{1}{2} \big[\openone_4 -\mathcal{G}(p^\mu)\big]  \label{eq:a}
\end{align}
and confirm that indeed they are projectors and furnish the  completeness relation
\begin{equation}
P_S^2 =P_S,\quad P_A^2 = P_A,\quad P_S+P_A = \openone_4. \label{eq:compl}
\end{equation}

Because $\mathcal{G}(p^\mu)$ is not Lorentz covariant its appearance in the spin sums 
violates Lorentz symmetry.
 In the past all efforts to circumvent this 
problem have failed and have given rise to a suggestion that the formalism can only be covariant under a subgroup of the Lorentz group suggested by Cohen and Glashow~\cite{Cohen:2006ky,Ahluwalia:2010zn}. \textit{However, we now report that there is a freedom in the definition of the dual. It allows a re-definition of the dual in such a way that the
Lorentz invariance of the orthonormality relations remains intact, 
but it restores the Lorentz covariance of the spin sums. }

To see this consider the following modification to the definition of the introduced dual
\begin{equation}
\dual{\lambda}^S_\alpha(p^\mu) \to \gdualn{\lambda}^S_\alpha(p^\mu) = \dual{\lambda}^S_\alpha(p^\mu) \mathcal{A},\quad
\dual{\lambda}^A_\alpha(p^\mu) \to \gdualn{\lambda}^A_\alpha(p^\mu) = \dual{\lambda}^A_\alpha(p^\mu) \mathcal{B}\label{eq:ab}
\end{equation}
with $\mathcal{A}$ and $\mathcal{B}$ constrained to have the 
following non-trivial properties: the $\lambda^S_\alpha(p^\mu)$ must be  eigenspinors of $\mathcal{A}$ with eigenvalue unity,  and similarly $\lambda^A_\alpha(p^\mu)$ must be  eigenspinors of $\mathcal{B}$ with eigenvalue unity\begin{equation}
\mathcal{A} \lambda^S_\alpha(p^\mu) = \lambda^S_\alpha(p^\mu),\quad
\mathcal{B} \lambda^A_\alpha(p^\mu) = \lambda^A_\alpha(p^\mu),\label{eq:4jan-a}\\
\end{equation}
and additionally $\mathcal{A}$ and $\mathcal{B}$ must be such that
\begin{equation}
\gdual{\lambda}^S_\alpha(p^\mu)\mathcal{A} \lambda^A_{\alpha^\prime}(p^\mu)=0,\quad \dual{\lambda}^A_\alpha(p^\mu)\mathcal{B} \lambda^S_{\alpha^\prime}(p^\mu) = 0.
\label{eq:4jan-b}
\end{equation}
 Under the new dual while the orthonormality relations (\ref{eq:zimpokJ9a})-(\ref{eq:zimpokJ9c}) would remain unaltered in form
\begin{align}
& \gdualn\lambda^S_\alpha(p^\mu) \lambda^S_{\alpha^\prime}(p^\mu)
 = 2 m \delta_{\alpha\alpha^\prime}\label{eq:zimpokJ9an}\\
&  \gdualn\lambda^A_\alpha(p^\mu) \lambda^A_{\alpha^\prime}(p^\mu)
 = - 2 m \delta_{\alpha\alpha^\prime} \label{eq:zimpokJ9bn} \\
 &  \gdualn\lambda^S_\alpha(p^\mu) \lambda^A_{\alpha^\prime}(p^\mu) = 0 =
 \gdualn\lambda^A_\alpha(p^\mu) \lambda^S_{\alpha^\prime}(p^\mu).\label{eq:zimpokJ9cn}
\end{align}
the same very  re-definition would alter the spin sums to
\begin{align}
\sum_{\alpha} \lambda^S_\alpha(p^\mu) \dualn \lambda^S_\alpha(p^\mu) & = 
m \big[\I_4 + \mathcal{G}(p^\mu) \big] \mathcal{A} \label{eq:sss-new}\\
\sum_{\alpha} \lambda^A_\alpha(p^\mu) \dualn \lambda^A_\alpha(p^\mu) & =
- m \big[\I_4 - \mathcal{G}(p^\mu)\big] \mathcal{B} \label{eq:ssa-new}
\end{align}
In what follow we will show that $\mathcal{A}$ and $\mathcal{B}$ exist that satisfy
the dual set of requirements encoded in (\ref{eq:4jan-a}) and (\ref{eq:4jan-b}) and at the same time find that a specific form of  $\mathcal{A}$ and $\mathcal{B}$ exists that renders the spin sums Lorentz invariant.

The spin sums determine the mass dimensionality of the quantum field that we will build from the here-constructed $\lambda(p^\mu)$ as its expansion coefficients. 
They enter the evaluation of the Feynman-Dyson propagator for the field to be introduced below.
For consistency with 
(\ref{eq:er-a1})-(\ref{eq:er-b2}) and (\ref{eq:skg}) this mass dimensionality must be one. And, this can only be achieved in the formalism we are developing if the spin sums are Lorentz invariant, and proportional to the identity.\footnote{So as to avoid confusion with the previous literature on the subject we make the following parenthetic remark:
If the spin sums are not Lorentz invariant then there is an internal inconsistency in the theory. It shows up in the form of a source term to be included in the kinematical equations of motion, and in non-locality -- in contradiction to the expectations based on equations (\ref{eq:er-a1})-(\ref{eq:er-b2}) and (\ref{eq:skg}). Strictly speaking, the usual arguments of determining the mass dimensionality of a quantum field are in the context of a local Lorentz covariant quantum field. While in a Lorentz breaking theory one may still invoke those arguments, as the literature on the subject did, from a strict theoretical point there is, and was, an unease in adoption of such a procedure. But that is the best one can do in such circumstances because simple dimensional arguments still hold.  The breakthrough on locality and Lorentz covariance reported here completely resolves what before was a grey area of using the arguments based on a local Lorentz covariant field in a context where the locality and Lorentz covariance were violated.
}

Thus, up to a constant (to be taken as $2$ to preserve orthonormality relations), $\mathcal{A}$ and  $\mathcal{B}$ must  be inverses of $\big[\I_4 + \mathcal{G}(p^\mu) \big]$ and 
$\big[\I_4 - \mathcal{G}(p^\mu) \big]$ respectively. But since 
the determinants of $\big[\I_4 \pm \mathcal{G}(p^\mu)\big]$ identically vanish we proceed in a manner akin to that of Penrose~\cite{PSP:2043984} and Lee~\cite{Lee:2014opa}, and with $\tau \in \Re$ we introduce 
a $\tau$   deformation of the spin sums (\ref{eq:sss-new})  and  (\ref{eq:ssa-new})  
\begin{align}
\sum_{\alpha} \lambda^S_\alpha(p^\mu) \dualn \lambda^S_\alpha(p^\mu) & = 
m \big[\I_4 + \tau\mathcal{G}(p^\mu) \big] \mathcal{A}\Big\vert_{\tau \to 1} \label{eq:sss-newnew}\\
\sum_{\alpha} \lambda^A_\alpha(p^\mu) \dualn \lambda^A_\alpha(p^\mu) & =
- m \big[\I_4 - \tau\mathcal{G}(p^\mu)\big] \mathcal{B}\Big\vert_{\tau \to 1}. \label{eq:ssa-newnew}
\end{align}
We will see that the $\tau \to 1$ limit is non pathological 
in the infinitesimal small neighbourhood of $\tau = 1$ in the sense we shall make explicit.
We choose $\mathcal{A}$ and $\mathcal{B}$ 
to be 
\begin{align}
& \mathcal{A} = 2 \big[I_4 + \tau \mathcal{G}(p^\mu)\big]^{-1} = 2 \left(\frac{\I_4 - \tau \mathcal{G}(p^\mu)}{1-\tau^2}\right) \\
& \mathcal{B} = 2 \big[I_4 - \tau \mathcal{G}(p^\mu)\big]^{-1} = 2 \left(\frac{\I_4 + \tau \mathcal{G}(p^\mu)}{1-\tau^2}\right)
\end{align}
Making use of the identity $\mathcal{G}^2(\phi) = \I_4$, Eqs.~(\ref{eq:sss-newnew}) and (\ref{eq:ssa-newnew}) simplify to:
\begin{align}
\sum_{\alpha} \lambda^S_\alpha(p^\mu) \dualn \lambda^S_\alpha(p^\mu) & = 
2 m \big[\I_4 + \tau\mathcal{G}(p^\mu) \big]   \left(\frac{\I_4 - \tau \mathcal{G}(p^\mu)}{1-\tau^2}\right)  \bigg\vert_{\tau \to 1} \nonumber \\
&=2 m \I_4 \left(\frac{1-\tau^2}{1-\tau^2}\right)\bigg\vert_{\tau\to 1}  =
 2m \I_4\label{eq:sss-newnew-a}\\
 \sum_{\alpha} \lambda^A_\alpha(p^\mu) \dualn \lambda^A_\alpha(p^\mu) & = 
2 m \big[\I_4 - \tau\mathcal{G}(p^\mu) \big]   \left(\frac{\I_4 + \tau \mathcal{G}(p^\mu)}{1-\tau^2}\right)  \bigg\vert_{\tau \to 1} \nonumber \\
&=2 m \I_4 \left(\frac{1-\tau^2}{1-\tau^2}\right)\bigg\vert_{\tau\to 1}  =
- 2m \I_4
 \label{eq:ssa-newnew-b}
\end{align}
We now return to the orthonormality relations. Since from the first equation in (\ref{eq:important}), $\mathcal{G}(p^\mu) \lambda^S(p^\mu) = 
\lambda^S(p^\mu)$ while $\mathcal{G}(p^\mu) \lambda^A(p^\mu) = -
\lambda^A(p^\mu)$, we have the result demanded by the requirement (\ref{eq:4jan-a})
\begin{align}
\mathcal{A} \lambda^S_\alpha(p^\mu)  = 2 \left(\frac{\I_4-\tau \mathcal{G}(p^\mu)}{1-\tau^2}\right) \lambda^S_\alpha(p^\mu)
= 2 \left(\frac{1-\tau}{1-\tau^2}\right) \lambda^S_\alpha(p^\mu) \nonumber\\
 = \left(\frac{2}{1+\tau}\right)\bigg\vert_{\tau\to 1}  \lambda^S_\alpha(p^\mu) = \lambda^S_\alpha(p^\mu)
\\
\mathcal{B} \lambda^A_\alpha(p^\mu) = 2 \left(\frac{\I_4+\tau \mathcal{G}(p^\mu)}{1-\tau^2}\right) \lambda^A_\alpha(p^\mu) 
= 2 \left(\frac{1-\tau}{1-\tau^2}\right) \lambda^A_\alpha(p^\mu) \nonumber\\
= \left(\frac{2}{1+\tau}\right)\bigg\vert_{\tau\to 1}  \lambda^A_\alpha(p^\mu) = \lambda^A_\alpha(p^\mu)
\end{align}
where  in the first two terms on the right hand side of each of the above equations
the ${\tau\to 1}$ limit has been suppressed. To examine the fulfilment of requirement (\ref{eq:4jan-b}) we note that
\begin{align}
\gdual{\lambda}^S_\alpha(p^\mu)\mathcal{A} \lambda^A_{\alpha^\prime}(p^\mu) = 
 2 \gdual{\lambda}^S_\alpha(p^\mu)\left(\frac{\I_4 - \tau \mathcal{G}(p^\mu)}{1-\tau^2}\right) \lambda^A_{\alpha^\prime}(p^\mu) \nonumber \\
=  2 \left(\frac{1}{1-\tau}\right)\bigg\vert_{\tau\to 1}
 \underbrace{\gdual{\lambda}^S_\alpha(p^\mu)\lambda^A_{\alpha^\prime}(p^\mu)}_{=\;0 ~\mbox{\small{(see eq. \ref{eq:zimpokJ9c})}}}
= \;0\\
\gdual{\lambda}^A_\alpha(p^\mu)\mathcal{B} \lambda^S_{\alpha^\prime}\alpha(p^\mu) = 
 2 \gdual{\lambda}^A_\alpha(p^\mu) \left(\frac{\I_4 + \tau \mathcal{G}(p^\mu)}{1-\tau^2}\right) \lambda^S_{\alpha^\prime}(p^\mu) \nonumber \\
 =  2 \left(\frac{1}{1-\tau}\right)\bigg\vert_{\tau\to 1}
 \underbrace{\gdual{\lambda}^A_\alpha(p^\mu)\lambda^S_{\alpha^\prime}(p^\mu)}
 _{=\;0 ~\mbox{\small{(see eq. \ref{eq:zimpokJ9c})}}}
= 0
\end{align}
where the  final equalities are to be understood as `in the infinitesimally  close neighbourhood of $\tau =1$, but not at $\tau=1$.' We will accept it as physically acceptable cost to be paid for the $\tau$ deformation forced upon us by the non-invertibility of $\big[\I_4 \pm \mathcal{G}(p^\mu)\big]$. With this caveat,
constraints (\ref{eq:4jan-a}) and (\ref{eq:4jan-b}) on $\mathcal{A}$ and 
$\mathcal{B}$ are satisfied resulting in the Lorentz invariant spin sums 
\begin{align}
&\sum_{\alpha} \lambda^S_\alpha(p^\mu) \dualn \lambda^S_\alpha(p^\mu) = 2m \I_4\label{eq:sss-newnew-a-new}\\
&\sum_{\alpha} \lambda^A_\alpha(p^\mu) \dualn \lambda^A_\alpha(p^\mu) = - 2m \I_4
 \label{eq:ssa-newnew-b-new}
\end{align}
without affecting the  Lorentz invariance of the  orthonormality relations (\ref{eq:zimpokJ9an})-(\ref{eq:zimpokJ9cn}).\footnote{
In Ref.~\cite{Rogerio:2016mxi}  Rogerio et al. have provide additional support for the new dual introduced here.}

Before introducing the new quantum field we return to the discussion surrounding Eq.~(\ref{eq:wrong}) and give the following as correct replacement for the `classical' Lagrangian density 
associated with the $\lambda(x)$
\begin{equation}
\mathfrak{L}(x) = \partial^\mu{\gdualn{\lambda}(x)}\,\partial_\mu {{\lambda(x)}} - m^2 {\gdualn{\lambda}}(x) \lambda(x).\label{eq:correct}
\end{equation}
The new dual has resolved all the problems including those encountered by Aitchison and Hey in Ref.~\cite[App. P]{Aitchison:2004cs}.

 \section{The new quantum field, its adjoint, Feynman-Dyson propagator, Lagrangian density, and zero point energy\label{sec:Elko-quantum-field}}

We now use the  $\lambda^S_\alpha(p^\mu)$ and $\lambda^A_\alpha(p^\mu)$ as expansion coefficients to define a new quantum field
\begin{align}
\mathfrak{f}(x) \stackrel{\mathrm{def}}{=} & \int \frac{\text{d}^3p}{(2\pi)^3}  \frac{1}{\sqrt{2 m E(\p)}}  \nonumber\\ &\times\sum_\alpha \Big[ a_\alpha(\p)\lambda^S_\alpha(\p) \exp(- i p_\mu x^\mu)
+\, b^\dagger_\alpha(\p)\lambda^A_\alpha(\p) \exp(i p_\mu x^\mu){\Big]}
\label{eq:newqf}
\end{align}
where we have taken the liberty to notationally replace the $\lambda(p^\mu)$ by $\lambda(\p)$ with the understanding that these eigenspinors are associated with a particle of mass $m$. The creation and annihilation operators satisfy Fermi statistics~\cite[Section~7]{Ahluwalia:2004ab} and~\cite[Section~4]{Ahluwalia:2015vea} 
\begin{align}
& \left\{a_\alpha(\p),a^\dagger_{\alpha^\prime}(\p^\prime)\right\} = \left(2 \pi \right)^3 \delta^3\hspace{-2pt}\left(\p-\p^\prime\right) \delta_{\alpha\alpha^\prime} \label{eq:a-ad}\\
& \left\{a_\alpha(\p),a_{\alpha^\prime}(\p^\prime)\right\} = 0,\quad \left\{a^\dagger_\alpha(\p),a^\dagger_{\alpha^\prime}(\p^\prime)\right\} =0\label{eq:aa-adad}
\end{align}
with similar anti-commutators for $b_\alpha(\p)$ and 
$b^\dagger_\alpha(\p)$.  The statistics is in fact dictated by causality, and this in turn also gives positivity to the Hamiltonian density~\cite{Ahluwalia:2015vea}.

To decipher the mass dimensionality of $\mathfrak{f}(x)$, we define the adjoint
 \begin{align}
\gdualn{\mathfrak{f}}(x) \stackrel{\mathrm{def}}{=}  & \int \frac{\text{d}^3p}{(2\pi)^3}   \frac{1}{ \sqrt{2 m E(\p)}} \nonumber\\ &\times
\sum_\alpha \Big[ a^\dagger_\alpha(\p)\gdualn{\lambda}^S_\alpha(\p) \exp( i p_\mu x^\mu)
 + b_\alpha(\p)\gdualn{\lambda}^A_\alpha(\p) \exp(-i p_\mu x^\mu){\Big]}.\label{eq:newadjoint}
\end{align}

 Using the above definitions of 
$\mathfrak{f}(x)$ and its adjoint $\gdualn{\mathfrak{f}}(x)$, 
and using the spin sums (\ref{eq:sss-newnew-a-new}) and  (\ref{eq:ssa-newnew-b-new}) in the intermediate calculations,
the amplitude for the particles described by the pair $\mathfrak{f}(x)$ and $\gdualn{\mathfrak{f}}(x)$ to go from spacetime point $x$ to $x^\prime$
is
\begin{align}
Q_{x\to x^\prime} &=  \xi  \left\langle\hspace{4pt}\left\vert \mathfrak{T} \left( \mathfrak{f}(x^\prime) \gdualn{f}(x)\right)\right\vert\hspace{4pt}\right\rangle \nonumber \\
& = 2 \xi i \int\frac{\text{d}^4 p}{(2 \pi)^4} 
{\e}^{- i p^\mu \left(x^{\prime}_\mu - x_\mu\right)} \left[  \frac{ \openone_4}{p_\mu p^\mu - m^2 + i \epsilon} \right] 
\end{align}
with $\epsilon = 0^+$.
The $\xi \in \C$ is determined so that, up to a global phase, when $Q_{x\to x^\prime} $  is integrated over all possible $x-x^\prime$ the result is unity, giving
\begin{equation}
\xi = \frac{i}{2} m^2
\end{equation}
With the consequence that 
\begin{equation}
Q_{x\to x^\prime} 
=  - m^2 \int\frac{\text{d}^4 p}{(2 \pi)^4} 
{\e}^{- i p^\mu \left(x^{\prime}_\mu - x_\mu\right)} \left[  \frac{ \openone_4}{p_\mu p^\mu - m^2 + i \epsilon} \right] 
\end{equation}

For calculational and conceptual details of this argument we refer our reader to section 6 of reference \cite{Ahluwalia:2004ab}. In the language of the just cited reference $\xi$ differs from $\varpi$ by a factor of half, $\xi = (1/2) \varpi $. The origin of this difference resides in the changes in the spin sums arising from additional constraint put on the Elko dual (see, equations (\ref{eq:sss-newnew-a-new})) and 
(\ref{eq:ssa-newnew-b-new}), the locality phases, and the change in pairing of the Elko with the creation and annihilation operators.

The Feynman-Dyson propagator is then defined to be proportional to $Q_{x\to x^\prime} $ in such a way that the proportionality constant is adjusted to make the Feynman-Dyson propagator  coincide with the Green function associated with the equation of motion for the field $\mathfrak{f}(x)$. The result of this analysis follows the details presented in \cite{Ahluwalia:2004ab}, but with the indicated changes, and reads
\begin{align}
S_{\mathrm{FD}}(x^\prime - x) & = -\frac{1}{m^2} \, Q_{x\to x^\prime}  \nonumber\\
& = \int\frac{\text{d}^4 p}{(2 \pi)^4} 
{\e}^{- i p^\mu \left(x^{\prime}_\mu - x_\mu\right)} \left[  \frac{ \openone_4}{p_\mu p^\mu - m^2 + i \epsilon} \right] \label{eq:FD-prop-b}
\end{align}
 It satisfies
\begin{equation}
\left(\partial_{\mu^\prime}\partial^{\mu^\prime} + m^2\, \I_4\right) S_{\mathrm{FD}}(x^\prime - x) =
- \delta^4\left(x^\prime - x\right)
\end{equation}
 This is an unexpected theoretical discovery that for a spin one half fermionic field based on Elko we do \underline{\textrm{not}} obtain
\begin{equation}
S_{\mathrm{FD}}(x^\prime - x) =
\int\frac{\text{d}^4 p}{(2 \pi)^4} 
{\e}^{- i p^\mu \left(x^{\prime}_\mu - x_\mu\right)} \left[  \frac{\gamma_\mu p^\mu + m \openone_4}{p_\mu p^\mu - m^2 + i \epsilon} \right]
\end{equation}
but instead (\ref{eq:FD-prop-b}).

As a consequence of (\ref{eq:FD-prop-b}), following the canonical discussion on the mass dimensionality of quantum fields given in Ref.~\cite[Section 12.1]{Weinberg:1995mt}
we find that mass dimension of the field $\mathfrak{f}(x)$ is  one
\begin{equation}
\mathfrak{D}_{\mathfrak{f}} = 1
\end{equation}
and not three-half, as is the case for the Dirac field.
Consequently, the free field Lagrangian density for the new field is
\begin{equation}
\mathfrak{L}_0(x) = \partial^\mu\gdualn{\mathfrak{f}}\,\partial_\mu {\mathfrak{f}}(x) - m^2 \gdualn{\mathfrak{f}}(x) \mathfrak{f}(x). \label{eq:fieldlagrangian}
\end{equation}

Following section 7 of reference~\cite{Ahluwalia:2004ab} we have re-analysed  the field energy  associated with the new fermionic field. We confirm that each of the four degrees of freedom associated with $\mathfrak{f}(x)$ carries a zero point energy  $= -(1/2) E(\p)$. The new zero point energy comes with a minus sign, and is infinite. For  an insightful  discussion on this point  we refer the reader to Matt Visser's recent e-print on the general subject~\cite{Visser:2016mtr}). Above the zero point energy, all the four degrees of freedom contribute the same energy $E(\p)$ to the field $\mathfrak{f}(x)$ for a given momentum $\p$.

\subsection{Interactions\label{sec:interactions}}

The mass dimension one and the power-counting arguments severely restrict interaction of the new fermions with the standard model particles, and confine them to the following 
\begin{equation}
\mathfrak{L}_{\text{int}}(x) = \lambda_1 \left(\gdualn{\mathfrak{f}}(x) \mathfrak{f}(x)\right)^2  + \lambda_2  \gdualn{\mathfrak{f}}(x) \mathfrak{f}(x) \,{\mathfrak{b}}^\dagger(x){\mathfrak{b}}(x)
\end{equation}
where $ \lambda_1$ and $\lambda_2$ are \emph{dimensionless} coupling constants. The first of these  is a  quartic self interaction, and the other is the interaction of the new field with any of the spin-zero bosonic fields $\mathfrak{b}(x)$, such as the Higgs~\cite{PhysRevLett.13.321,PhysRevLett.13.508,PhysRevLett.13.585}. For the mass dimension three-half Dirac field, similar interactions are suppressed, respectively, by two and one powers of the unification/Planck scale.

This opens up the possibility that the new field provides a natural self-interacting dark matter candidate. The darkness arises from two related facts: one, due to mass dimension mismatch of the standard model fermions and the mass dimension one fermions the latter cannot enter the standard model doublets; and two, the formalism for mass dimension one fermions does not support the standard model local gauge interactions.

Beyond these dimension-four interactions one may also introduce the following Yukawa couplings of dimension \textit{three and half} (that is, $7/2$) with neutrinos \cite{ArnabDasgupta} 
\begin{equation}
\mathfrak{L}_{\text{int}}(x) = \ell_1 \,
\phi(x) \,\overline{\nu}(x)\mathfrak{f}(x) + 
 \ell_2 \,\phi(x)  \gdualn{\mathfrak{f}} (x) {\nu}(x)\label{eq:Yukawa}
\end{equation}
where $ \ell_1$ and $\ell_2$ are dimensionfull coupling constants, and $\nu(x)$ is a Dirac or a Majorana field. These may be used to violate lepton number.

\subsection{Locality structure of the new field\label{sec:locality}}

Now that we have $\mathfrak{L}_0(x)$ we can calculate the momentum conjugate to $\mathfrak{f}(x)$
\begin{equation}
\mathfrak{p}(x) = \frac{\partial \mathfrak{L}_0(x)}
{\partial {\dot{\mathfrak{f}}(x)}} = \frac{\partial}{\partial t}\gdualn{\mathfrak{f}}(x).
\end{equation}
To establish that the new field is local we calculate the standard equal-time anti-commutators. The first of the three anti-commutators we calculate is the `$\mathfrak{f}$-$\mathfrak{p}$' anti-commutator 
\begin{equation}
\left\{ \mathfrak{f}(t,\x),\;\mathfrak{p}(t,\x^\prime) \right\}.
\end{equation}
It evaluates to
\begin{equation}
i \int \frac{\text{d}^3 p}{(2 \pi)^3} \frac{\e^{i\p\cdot(\x-\x^\prime)}}{4 m}
\sum_\alpha\left[ \lambda^S_\alpha(\p) \gdualn{\lambda}^S_\alpha(\p)  - 
 \lambda^A_\alpha(-\p) \gdualn{\lambda}^A_\alpha(-\p) \right].\label{eq:zimpok-loc-1}
\end{equation}
The spin sums are independent of $p^\mu$, therefore (\ref{eq:ssa-newnew-b-new}) gives
\begin{equation}
\sum_\alpha  \lambda^A_\alpha(-\p) \gdualn{\lambda}^A_\alpha(-\p)  = -
2 m \I_4.
\end{equation}
with the result 
\begin{equation}
\sum_\alpha\left[ \lambda^S_\alpha(\p) \gdualn{\lambda}^S_\alpha(\p)  - 
 \lambda^A_\alpha(-\p) \gdualn{\lambda}^A_\alpha(-\p) \right] = 4 m \I_4
\end{equation}
Consequently we have (without the non-locality inducing extra term of all previous papers on the subject~\cite{Ahluwalia:2004sz,Ahluwalia:2004ab,Ahluwalia:2008xi,Ahluwalia:2009rh,Ahluwalia:2010zn}) 
\begin{equation}
\left\{ \mathfrak{f}(t,\x),\;\mathfrak{p}(t,\x^\prime) \right\} = i \delta^3\left(\x-\x^\prime\right) \openone_4 .\label{eq:lac-1}
\end{equation}
A still simpler calculation shows that the remaining two, that is, `$\mathfrak{f}$-$\mathfrak{f}$' and  `$\mathfrak{p}$-$\mathfrak{p}$', equal time anti-commutators 
vanish
\begin{equation}
\left\{ \mathfrak{f}(t,\x),\;\mathfrak{f}(t,\x^\prime) \right\} = 0, \quad 
\left\{ \mathfrak{p}(t,\x),\;\mathfrak{p}(t,\x^\prime) \right\} = 0.\label{eq:lac-2and3}
\end{equation}
The field $\mathfrak{f}(x)$ is thus local in the sense of Schwinger~\cite[Sec. II, Eqs. 2.82]{PhysRev.82.914}. It is a much stronger condition of locality than that adopted by Schwartz~\cite[Sec. 24.4]{Schwartz:2013pla}.

\subsection{Majorana-isation of the new field\label{sec:Majorana-isation}}

Even though field $\mathfrak{f}(x)$ is uncharged under local U(1) supported by the Dirac fields of the standard model of high energy physics, it may carry a charge under a different  local U(1) gauge symmetry such as the one suggested in the discussion around (\ref{eq:mathfraka}). This gives rise to the  possibility of having a fundamentally neutral field in the sense of Majorana~\cite{Majorana:1937vz}
\begin{align}
\mathfrak{m}(x) &=   \int \frac{\text{d}^3p}{(2\pi)^3}  \frac{1}{ \sqrt{2 m E(\p)}} \nonumber\\ &\times \sum_\alpha \Big[ a_\alpha(\p)\lambda^S(\p) \exp(- i p_\mu x^\mu)
+ \, a^\dagger_\alpha(\p)\gdualn\lambda^A(\p) \exp(i p_\mu x^\mu){\Big]} 
\end{align}
with momentum conjugate 
\begin{equation}
\mathfrak{q} = \frac{\partial}{\partial t}\gdualn{\mathfrak{m}}(x).
\end{equation}
The calculation for the `$\mathfrak{m}$-$\mathfrak{q}$' equal time anti-commutators goes through exactly as before and one gets
\begin{equation}
\left\{ \mathfrak{m}(t,\x),\;\mathfrak{q}(t,\x^\prime) \right\} = i \delta^3\left(\x-\x^\prime\right) \openone_4 .
\end{equation}
The calculation of the remaining two anti-commutators requires knowledge of the following 
`twisted' spin sums 
\begin{align}
&\sum_\alpha\left[ \lambda^S_\alpha(\p)\left[ {\lambda}^A_\alpha(\p)\right]^T  + 
 \lambda^A_\alpha(-\p) \left[{\lambda}^S_\alpha(-\p)\right]^T  \right]\\
 & \sum_\alpha\left[ \left[\gdualn{\lambda}^S_\alpha(\p)\right]^T {\gdualn{\lambda}}^A_\alpha(\p)  + 
 \left[\gdualn{\lambda}^A_\alpha(-\p)\right]^T \gdualn{\lambda}^S_\alpha(-\p)\right].
\end{align}
One finds that each of these vanishes. With this result at hand, we immediately decipher vanishing of the `$\mathfrak{m}$-$\mathfrak{m}$' and  `$\mathfrak{q}$-$\mathfrak{q}$', equal time anti-commutators 
\begin{equation}
\left\{ \mathfrak{m}(t,\x),\;\mathfrak{m}(t,\x^\prime) \right\} = 0, \quad 
\left\{ \mathfrak{q}(t,\x),\;\mathfrak{q}(t,\x^\prime) \right\} = 0.
\end{equation}
The field $\mathfrak{m}(x)$, like $\mathfrak{f}(x)$, is thus local in the sense of Schwinger~\cite[Sec. II, Eqs. 2.82]{PhysRev.82.914}.

\section{A guide to the existing literature on Elko and mass dimension one fermions\label{sec:literature}}

As soon as the first papers introducing Elko and mass dimension one fermions were 
published~\cite{Ahluwalia:2004sz,Ahluwalia:2004ab} da Rocha and Rodrigues Jr. noted that 
Elko belong to class 5 spinors~\cite{daRocha:2005ti}  in the Lounesto classification~\cite[Chapter 12]{Lounesto:2001zz}.  The possibility that the new fermions may be a dark matter candidate was first noted, besides our papers, in~\cite{Lazarides:2006jw} of George Lazarides. The possibility that in the early universe fermions different than Diarc/Majorana may exist was briefly noted by Guendelman and Kaganovich~\cite{Guendelman:2006ji}. A coupling identical in form to the Yukawa coupling 
(\ref{eq:Yukawa}) for mass dimension one fermions appears there first.

Christian B\"ohmer was the first to note that the spin angular momentum tensor associated with Elko cannot be entirely expressed as an axial torsion vector~\cite{Boehmer:2006qq}. He emphasised that this important difference from the Dirac spinors arises due to different helicity structures of the Elko and Dirac spinors. His groundbreaking paper also put forward a tiny coupling of Elko spinors to Yang-Mills fields and discussed its implications for consistently coupling massive spin one field to the Einstein-Cartan theory. Restricting to the  Einstein-Elko system~ ~\cite{Boehmer:2006qq}  he constructed analytical ghost Elko solutions with the property of a vanishing energy-momentum tensor.
This was done to make the analytical calculations possible\footnote{This assumption was later placed on a more natural footing by~\cite{Chang:2015ufa} \textit{et al.}} and he showed that, ``the Elko spinors are not only prime dark matter candidates but also prime candidates for inflation.'' With his collaborators 
B\"ohmer has placed  Elko cosmology on a firm footing with an eye on the available 
data. We refer the reader to references~\cite{Boehmer:2007ut,Boehmer:2008rz,Boehmer:2008ah,Boehmer:2009aw,Boehmer:2010ma} for details. While building Elko cosmology he has coined the term ``dark spinors'' for Elko.

The group of Julio Hoff da Silva and Saulo Pereira, focusing on exact analytical solutions, have taken Elko cosmology significantly beyond 
B\"ohmer's initial pioneering efforts. We refer the reader to their 
publications~\cite{daSilva:2014kfa,Pereira:2014wta,S:2014dja,Pereira:2014pqa}.
Concurrently 	extending the work of B\"ohmer, Gredat and Shankaranarayanan have considered an Elko-condensate driven inflation and shown that it is favoured by existing observational data~\cite{Gredat:2008qf}. This work has been followed by Basak and Shankaranarayanan to prove that, 
``Elko driven inflation can generate growing vector modes even in the first order." This allows them to generate vorticity during inflation to produce primordial magnetic field~\cite{Basak:2014qea}. In a related publication Basak et al. show that Elko cosmology provides two sets of attractor points. These correspond to slow and fast-roll inflation. The latter being unique to Elko~\cite{Basak:2012sn}. For earlier contribution to Elko cosmology from this group we refer to the reader to references~\cite{Shankaranarayanan:2010st,Shankaranarayanan:2009sz}. The cosmological coincidence problem in the context of Elko is discussed by Hao Wei in reference~\cite{Wei:2010ad}, while one of the early papers on stability of de Sitter solution in the context of Elko is reference~\cite{Chee:2010ju}.

Elko cosmology has gained a significant and independent  boost through a recent study of phantom dark-energy Elko/dark-spinors  undertaken by Yu-Chiao Chang \textit{et al.} In the context of Einstein-Cartan theory, it resolves a host of problems with phantom dark energy models and predicts a final  de Sitter phase for our universe at late time with or without dark matter~\cite{Chang:2015ufa}. Their work not only makes Elko and mass dimension one fermions more physically viable but it also lends concrete physicality to torsion as an important possible element of reality.
	
The Brazilian-Italian group of Rold\~ao da Rocha, Hoff da Silva,  Bueno Rogerio, Cavalcanti,  
Bonora, Fabbri, Silva-Neto, Bernardini, J. G. Pereira, and Coronado Villalobos, besides examining such 
important topics as Hawking radiation of mass dimension one fermions~\cite{daRocha:2014dla}, continue to develop mathematical physics underlying  
Elko~\cite{daRocha:2007sd,daRocha:2007pz,daRocha:2008we,HoffdaSilva:2009is,daRocha:2011xb,daRocha:2011yr,Bernardini:2012sc,daSilva:2012wp,daRocha:2013qhu,Cavalcanti:2014uta,daRocha:2014dla,Bonora:2015ppa,daRocha:2016bil,Rogerio:2016grn,daSilva:2016htz,HoffdaSilva:2017vic,Neto:2017vgx}.
Of these we draw particular attention to mass-dimension transmuting operators considered in
~\cite{daRocha:2007pz,HoffdaSilva:2009is}. It would  help define a new symmetry between the Dirac field and the field associated with mass dimension one fermions if mass-dimension transmuting operators could be placed on a rigorous footing after incorporating locality and Lorentz covariance reported in this communication.\footnote{The need for a mass-dimension transmuting symmetry was first suggested by the present author to Rold\~ao da Rocha several years ago.} 

The problem of dark energy, and having a first-principle candidate for it, is one of the most challenging problem in Physics. In 2008,
Max Chaves and Doug Singleton suggested that mass dimension one fermions of spin one half  may have a possible connection with mass-dimension-one vector particles with fermionic statistics~\cite{Chaves:2008gd}. It may be worth examining if a new fundamental symmetry may be constructed that relates the works of~\cite{daRocha:2007pz,HoffdaSilva:2009is} with those of Chaves and Singleton.

Localisation of Elko in the brane has been considered in references~\cite{Liu:2011nb,Jardim:2014xla}. Elko in the presence of torsion has been a subject of several insightful papers by Luca Fabbri. We refer the reader to these and related publications~\cite{Fabbri:2009ka,Fabbri:2010qv,Fabbri:2010ws,Fabbri:2010va,Fabbri:2011mi,Fabbri:2012yg,Fabbri:2014foa}. Cosmological solutions of 5D Einstein equations with Elko condensates were obtained by Tae Hoon Lee where it was found that there exist exponentially expanding cosmological solution even in the absence of a cosmological constant~\cite{Lee:2012zze}.

All the works discussed so far remain essentially unchanged with the new developments reported here. In view of the results on locality and Lorentz covariance reported here it is important to revisit the analysis and claims of~\cite{Basak:2011wp} and also those calculations that use full apparatus of the theory of quantum fields, and not merely Elko.
In the same thread, given the interest in mass dimension one fermions a  number of $S$-matrix  calculations were done and published~\cite{Dias:2010aa,Lee:2015sqj,Alves:2014qua,Alves:2014kta,Agarwal:2014oaa,Lee:2015jpa}. These need to be revisited also.

Recently Saulo Pereira and R. C. Lima have claimed that 
an asymptotically expanding universe creates
low-mass mass dimension one fermions much more copiously than Dirac fermions (of the same mass)~\cite{Pereira:2016eez}. If their preliminary results remain essentially unaffected by the new results presented here it would significantly help us to develop a first-principle cosmology based on Elko and mass dimension one fermions. 

\section{Conclusion\label{sec:conclusion}}

With this communication we have achieved a breakthrough on locality and Lorentz covariance of mass dimension one fermions of spin one half. The field $\mathfrak{f}(x)$ depends on the eigenspinors of the charge conjugation operator in precisely the same manner as the Dirac field is determined by the eigenspinors of the parity operator. We succeeded in evading the expectations based on the no go theorems contained in the work of Wigner, Lee and Wick, and Weinberg~\cite{Wigner:1962ep,PhysRev.133.B1318,Weinberg:1995mt}  by carefully examining the underlying structure that lies underneath duals and adjoints. And by constructing a new dual for the eigenspinors of the charge conjugations operator and using it to define a new spin one half field, and its adjoint. In constructing the new field we were  intricately helped by the work of Weinberg on pairing of the expansion coefficients with the creation and annihilation operators in a very specific manner.

We argued that the new fermionic field  has extremely limited interactions with the standard model matter and gauge fields and that it provides  first-principle dark matter fields. Its dimension four couplings are restricted to a quartic self interaction, and to Higgs. It also supports a Yukawa-like coupling of dimension three and one half  with neutrinos.

Finally, we brought in focus that a first-principle cosmology based on Elko and mass dimension one fermions has taken birth. In it the standard model particles, dark matter, inflation, and possibly dark energy are merged in one whole where spacetime symmetries and certain discrete symmetries play a crucial role.

\subsection*{Acknowledgment}

 I thank Llohann 
Speran\c{c}a for discussions in the initial stages of this manuscript at Unicamp (S\~ao Paulo).
The breakthrough on the Lorentz symmetry and locality presented here began in late 2015 during a  three-month long visit to the Inter-University Centre for Astronomy and Astrophysics where I gave a series of lectures on mass dimension one fermions. For their insightful questions and the ensuing discussions, I thank  the participants of those lectures and in particular Sourav Bhattachaya, Sumanta Chakraborty, Swagat Mishra, Karthik Rajeev, and Krishna Parattu.
Raghu Rangarajan (Physical Research Laboratory) carefully read the entire first draft of the manuscript and provided many insightful suggestions. I am grateful to him for
his generosity and for engaging in long insightful discussions. The calculations that led to the reported results were done at Centre for the Studies of the Glass Bead Game and Physical Research Laboratory.  I thank Indian Institute of Technology Gauhati for providing me an ambiance of quiet and beauty to write the final draft of this manuscript.

\vspace{21pt}


\providecommand{\href}[2]{#2}\begingroup\raggedright\endgroup

\end{document}